\documentclass[%
 reprint,
 amsmath,amssymb,
 aps,
 longbibliography,
]{revtex4-1}

\usepackage{graphicx}
\usepackage{dcolumn}
\usepackage{bm}
\usepackage{hyperref}
\usepackage[normalem]{ulem}
\hypersetup{
    colorlinks=true,
    linkcolor=blue,
    filecolor=blue,      
    urlcolor=blue,
    citecolor=blue,
}

\begin{document}
\title{Rank-3 Moment Closures in General Relativistic Neutrino Transport}
\author{Sherwood Richers}
\affiliation{Department of Physics, University of California, Berkeley, CA 94720, USA}

\begin{abstract}
Many modern simulations of accretion disks use moment-based methods for radiation transport to determine the thermal evolution of the disk and the properties of the ejected matter. The popular M1 scheme evolves the rank-0 and rank-1 moments requires an analytic approximation for the rank-2 and higher tensors. We present the open-source Monte Carlo steady-state general-relativistic neutrino transport code {\tt SedonuGR}, which we use to asses fundamental analytic closure assumptions, quantify proposed closure errors, and test an extension of the maximum-entropy Fermi-Dirac (MEFD) closure to the rank-3 moment. We demonstrate that the fundamental assumptions employed in all analytic closures are strongly violated. This violation is most evident at the interface between the equatorial disk and the evacuated polar regions. Finally, we calculate the neutrino momentum and energy deposition rate from neutrino pair annihilation, and demonstrate that a moment-based annihilation power calculation is accurate to at most $\sim20\%$ if the rank-2 and higher moments are neglected. Out of a selection of 8 closures in the literature, we demonstrate that different closures reproduce different aspects of the radiation field (pressure tensor, rank-3 tensor, pair annihilation rate), though the MEFD, Levermore, and Janka 2 closures are all reasonable. The extra information from the neutrino degeneracy used in the MEFD closure is unable to account for the diversity in the rank-2 and rank-3 moments.
\end{abstract}

\maketitle

\section{Introduction}
\label{sec:introduction}
The landmark 2017 multimessenger detection of a binary neutron star merger \cite{Abbott2017} confirmed an evolutionary channel of compact objects that had been theoretically speculated for decades \cite{Lattimer1974,Li1998a,Rosswog2003}. In addition, a handful of other events with less significance but consistent with binary neutron star mergers have been observed without gravitational waves \cite{Tanvir2013a,Yang2015}, without an electromagnetic counterpart \cite{Abbott2020}, or as a possible neutron star-black hole merger \cite{Abbott2020a}. The promise of future multimessenger detections \cite{Pol2019} provides the possibility of constraining the nuclear equation of state \cite{Margalit2019}, confirming the origin of heavy elements in the universe \cite{Cote2018}, and understanding the nature of the engine that drives gamma-ray bursts \cite{Abbott2019}.

Following years of purely theoretical work and a flurry of papers following GW170817, a standard model of the merger dynamics has emerged (see \cite{Shibata2019,Radice2020,Metzger2020} for recent reviews), though this model is incomplete and numerical simulations still exhibit a good deal of uncertainty \cite{Foucart2016a,Foucart2018a,Radice2018a,Foucart2019}. During the inspiral and merger of a pair of neutron stars, a tidal ejecta is launched in the equatorial regions with a very low electron fraction that leads to a red transient due to efficient production of heavy elements via the rapid neutron capture process. There remains a central compact object that can be a stable neutron star, temporarily stable hypermassive neutron star, or black hole, depending on the details of the merging system and currently uncertain details about the nuclear equation of state (e.g., \cite{Radice2018b}). Around the central object is a hot and dense accretion disk, the mass of which depends on all of the same factors as well \cite{Radice2018a}. Emission of neutrinos allows the disk to cool and raise the electron fraction toward 0.5. The accretion disk itself also launches a significant amount of matter due to a combination of viscous or MHD stresses and neutrino heating\cite{Miller2019b,Fernandez2020}, and may launch a polar jet via MHD stresses \cite{Blandford1977,Ruiz2020} or neutrino pair annihilation \cite{Eichler1989,Just2016,Perego2017,Fujibayashi2017}. In any case, neutrinos play a significant role in transporting energy and lepton number, which can determine the amount and composition of the ejecta, which then determines the electromagnetic observables like the brightness, color, and duration of a kilonova \cite{Metzger2014,Radice2018a}. 

The ejected matter enshrouds the inner mass ejection engine, obscuring our view of the details of the accretion disk dynamics. Because of this, we resort to theoretical and numerical models to interpret and predict observables, but the combination of turbulent hydrodynamics, strong spacetime curvature, an uncertain nuclear equation of state, and strong asymmetries make understanding the dynamics of the radiation a challenging problem. The general relativistic quantum kinetic equations must be solved to fully understand the neutrino dynamics \cite{Vlasenko2014,Volpe2015,Blaschke2016}. Though efforts are being made to understand the role of neutrino flavor changes in neutron star mergers (e.g., \cite{Wu2017c}), state of the art dynamical models generally ignore this wrinkle and solve or approximate the Boltzmann equation. The most exact methods currently employed are Monte Carlo methods (e.g. \cite{Miller2019b}). Such accurate methods are currently prohibitively expensive in extremely dense regions like the interior of a hypermassive neutron star and for carrying out simulations beyond a couple hundred milliseconds. Other exact neutrino transport methods have been used in the context of multidimensional core-collapse supernova simulations, but have not yet been tested in such a relativistic environment as neutron star mergers (e.g., \cite{Sumiyoshi2012,Nagakura2014,Nagakura2017a}).

The most efficient approximate treatments of neutrinos in merger disks employ the leakage scheme \cite{Metzger2014,Siegel2018,Fernandez2020}, the advanced spectral leakage scheme \cite{Gizzi2019}, or a combination of leakage and moment treatments \cite{Sekiguchi2015,Radice2016a,Fujibayashi2017}. Two-moment methods \cite{Thorne1981,Shibata2011,Cardall2013} are a more sophisticated and very popular approximation to the Boltzmann equations in all types of astrophysical accretion disks, including protoplanetary disks \cite{Fuksman2020}, active galactic nuclei \cite{Gonzalez2007,Sadowski2013,Skinner2013,Fragile2014,Jiang2014,Stone2020}, and compact object merger remnants \cite{Foucart2018a,Fujibayashi2020}. However, these methods still require some scheme for estimating higher-rank moments in order to complete the system of equations. While methods exist to evaluate these higher moments using the method of short characteristics (e.g., \cite{Davis2012,Jiang2019,Weih2020}) or potentially Monte Carlo transport \cite{Foucart2018}, it is much more efficient to use an analytic closure method to determine the higher moments as a function only of local quantities \cite{Thorne1981,Shibata2011,Cardall2013}. The analytic closure is implemented by expressing the pressure tensor as a function of the energy density and flux, most often according to the Levermore closure \cite{Levermore1984}.

One of the exciting features of moment-based methods is that if the closure is exact the evolution equations for the evolved moments are exact. There has been a great deal of effort to find a closure relation that approaches this ideal, though largely restricted to one dimension (see \cite{Cernohorsky1994,Murchikova2017} for summaries). \cite{Foucart2018a} analyze the difference between radiation fields evolved using a dynamical gray moment method and a Monte Carlo method evolved in parallel but without feedback to the fluid. They showed that moment methods fail to accurately reproduce neutrino average energies in the equatorial region, neutrino densities in the polar region, and neutrino pair annihilation rates, likely affecting mass outflow from the disk. The question naturally remains, however, of whether it is possible to invent a local closure that is adequately realistic. \cite{Iwakami2020} demonstrated that in multidimensional core-collapse supernova simulations, the pressure tensor can be significantly misaligned with the flux vector, violating a fundamental assumption that goes into the closure approximation. In addition, although the closure for the rank-2 moment (pressure tensor) is often discussed, there has yet been no analysis of the quality of the rank-3 moment closure needed for spectral moment schemes. In this paper we extend the maximum entropy Fermi-Dirac (MEFD) closure to include the rank-3 moments, investigate how a realistic radiation field breaks the fundamental assumptions used by any local analytic closure, compare several closures suggested in the literature, and identify where the choice of closure has the largest impact.

The paper is organized as follows. In Section~\ref{sec:SedonuGR} we introduce our upgraded general-relativistic Monte Carlo radiation transport code. We discuss the analytic closure approximation in Section~\ref{sec:closure} and proceed to derive the maximum-entropy Fermi-Dirac closure for the rank-3 moment tensor. We calculate a steady-state neutrino radiation field on a single snapshot of a neutron star merger simulation in Section~\ref{sec:results}, carefully inspect the validity of the assumptions that go into the analytic closures, quantify errors from several closures in the literature, and assess the impact that the closure choice has on the neutrino pair annihilation rate. Finally, we conclude in Section~\ref{sec:conclusions} with a discussion of the possibility of improving the closure in a general way. We focus our attention on the neutron star merger environment, but many of the conclusions in this paper are relevant to any system where moment-based radiation transport algorithms are used.

\section{Discrete General Relativistic Monte Carlo Transport}
\label{sec:SedonuGR}
\texttt{SedonuGR} is a time-independent general-relativistic (GR) neutrino radiation transport code that operates in zero (spatially homogeneous) to three dimensional systems. This is a heavily modified and upgraded version of the special-relativistic code \texttt{Sedonu} neutrino transport code \cite{Richers2015, Richers2017a} and is publicly available \cite{SedonuGR}\footnote{\url{github.com/srichers/SedonuGR}}. In this section we will describe the code in detail, and several code tests are presented in Appendix~\ref{app:code_tests}.

The general relativistic Boltzmann equation describes the evolution of the neutrino distribution $f_\epsilon$ as \cite{Thorne1981}
\begin{equation}
    \frac{d x^\alpha}{d\tau}\frac{\partial f_\epsilon}{\partial x^\alpha} + \frac{d k^i}{d\tau}\frac{\partial f_{{\epsilon}}}{\partial k^i} = -k^\alpha u_\alpha S\,\,,
\end{equation}
where $d\tau$ is an interval of time in the rest frame of the background fluid, $x^\alpha$ is the neutrino position, $k^\alpha$ is the neutrino four-momentum (in units of energy), $u^\alpha$ is the four-velocity, and $S$ is a source term that accounts for collisions. The current goal of {\tt SedonuGR} is to solve the time-independent version of this equation, namely under the assumption that $\partial_t f_\epsilon=0$. We solve this equation via Monte Carlo transport \cite{Haghighat2015} by discretizing the distribution function into a finite number of neutrino packets, each of which undergo random emission, propagation, and scattering just as real individual neutrinos would. This section is an exposition of the details of the method, and the reader interested in the results of the calculations can jump to Section~\ref{sec:results}.

\subsection{Coordinates}
\label{sec:coordinates}
The Monte Carlo neutrino packets use the standard 3+1 Cartesian metric with the (-+++) sign convention describing coordinates in the \textit{lab frame}:
\begin{equation}
  \begin{aligned}
    g_{\alpha\beta} &=
    \begin{bmatrix}
      -\alpha^2 + \beta_\alpha\beta^\alpha & \beta_i \\
      \beta_j & \gamma_{ij}
    \end{bmatrix}\,\,,\\
    g^{\alpha\beta} &=
    \begin{bmatrix}
      -\alpha^{-2} & \alpha^{-2}\beta^i \\
      \alpha^{-2}\beta^j & \gamma^{ij} - \alpha^{-2}\beta^i\beta^j
    \end{bmatrix}\,\,.\\
  \end{aligned}
\end{equation}
where $\alpha$ is the lapse, $\beta^\alpha$ is the shift vector, and $\gamma_{ij}$ is the three-metric. A location is specified with a four-coordinate $x^\mu$ and a neutrino momentum is specified with a wavevector $k^\alpha$. All particle motion is done in the lab frame in 3D Cartesian coordinates. The shift vector can be chosen freely \cite{Baumgarte2013}, affecting how the spacetime is evolved. Since we are not allowing the spacetime to evolve, we must choose the shift vector in a way that is consistent with the volume of a spatial cell not to changing in time. That is, we must choose $\beta^i$ such that the extrinsic curvature vanishes, which is most simply done by choosing $\beta^i=0$.

With the metric quantities and the three-velocity in these coordinates given by a simulation snapshot, we reconstruct the dimensionless four-velocity as
\begin{equation}
\begin{aligned}
u^t &= \frac{W}{\alpha}\,\,, \\
u^i &= W\left(\frac{v^i}{c} - \frac{\beta^i}{\alpha}\right)\,\,.
\end{aligned}
\end{equation}
where $W = 1/\sqrt{1-\gamma_{ij}v^i v^j}$.

In addition to the \textit{lab} frame, we also require the ability to define a local comoving orthonormal \textit{tetrad} defined by orthonormal basis vectors $e_{(\mu)}^\alpha$. This frame is used when performing neutrino-fluid interactions or aggregating the radiation field. When constructing a tetrad at a particular location, the timelike basis vector is the fluid four-velocity at that location ($e^\alpha_{(t)} = u^\alpha$), yielding a comoving coordinate system. Following \cite{Dolence2009}, we provide a trial vector, subtract off components not normal to each of the previously determined basis vectors, and renormalize the vector to make it have a magnitude of unity. To make a trial vector $e_{(\mathrm{trial})}^\alpha$ orthogonal to another vector $l^\alpha$, we set $e_\mathrm{trial}^\alpha \gets e_\mathrm{trial}^\alpha - e_\mathrm{trial}^\alpha l_\alpha / l^\alpha l_\alpha$. To normalize, we set $e_\mathrm{trial}^\alpha \gets e_\mathrm{trial}^\alpha / \sqrt{e_\mathrm{trial}^\alpha e^\mathrm{trial}_\alpha}$. Once the basis vectors are established, we can transform a four-vector into the comoving tetrad basis and back out via
\begin{equation}
\label{eq:tet_transform}
\begin{aligned}
k^\alpha_\mathrm{tet} &= k^\mu e_{(\alpha)}^{\nu}g_{\mu\nu}\,\,,\\
k^\alpha &= k^\mu_\mathrm{tet} e_{(\mu)}^{\alpha}\,\,.
\end{aligned}
\end{equation}

All Monte Carlo packets use these coordinates independent of the structure of the underlying data. The background fluid and metric data are stored in a grid unused by the neutrino packets themselves, and a grid class specific to each type of geometry (homogeneous, 1D spherical, 2D spherical, 3D Cartesian) knows how to interpolate all quantities and derivatives (Section~\ref{sec:interpolation}) to the MC particle position $x^\alpha$ and momentum $k^\alpha$ in these Cartesian coordinates. Below, we describe how the neutrino packets interface with background data stored in three-dimensional Cartesian and one-dimensional spherical-polar grids.

In addition to the spatial grid, we also use a discrete energy grid with NG bins, centered at energy $\omega_i$ and with grid boundaries at $\omega_{i-1/2}$ and $\omega_{i+1/2}$, where $i$ ranges from 1 to NG. The neutrino energies are defined as $\omega=-k^\alpha u_\alpha$, or simply the energy in the comoving frame.


\subsubsection{1D Spherical-Polar Background}
We use the 1D spherical geometry in the code tests in Appendix~\ref{app:code_tests}. In spherical symmetry, the metric in general is $ds^2 = -\alpha^2 dt^2 + X^2 dr^2 + r^2d\Omega^2$, where the spherical coordinates can be expressed in terms of the Cartesian coordinates as $r=\sqrt{x^i x^i}$, $\theta=\cos^{-1}(x^z/r)$, and $\phi=\tan^{-1}(x^y/x^x)$.

We store and interpolate $\alpha$, $X$, and $v^r$ to a given neutrino position in the spherical grid. We can then reconstruct the three-dimensional three-velocity components and metric as
\begin{equation}
\begin{aligned}
v^i &= v^r \frac{x^i}{r}\,\,,\\
\gamma_{ij} &= \frac{x^i x^j}{r^2} (X^2-1) + \delta_{ij}\,\,.
\end{aligned}
\end{equation}
Given the radial derivatives of $\alpha$ and $X$, we can also reconstruct the Christoffel symbols as
\begin{equation}
\begin{aligned}
&\Gamma^t_{\mu\mu} = \Gamma^t_{ij} = 0\,\,, \\
&\Gamma^t_{ti} = \frac{x^i}{r} \frac{\partial_r \alpha}{\alpha} \,\,,\\
&\Gamma^a_{tt} = \frac{x^a}{r} \frac{\alpha \partial_r \alpha}{X^2}\,\,,\\
&\begin{aligned} \Gamma^a_{ij} = 
\frac{x^a}{(Xr)^2} &\left[\frac{x^i x^j}{r^2}(1-X^2 + rX \partial_r X)\right. \\
&\left.- \delta_{ij} (1-X^2)\right]\,\,.
\end{aligned}
\end{aligned}
\end{equation}
In the above, radial derivatives are computed via finite differencing between the nearest neighbors, and time derivatives are assumed to be zero.

When constructing a comoving orthonormal tetrad, our trial vectors are $\{xz, yz, -\widetilde{r}^2, 0\}$, $\{-y, x, 0, 0\}$, and $\{x, y, z, 0\}$, where $\widetilde{r}^2 = x^2+y^2$. These correspond to the $\theta$, $\phi$, and $r$ directions, respectively. 

\subsubsection{3D Cartesian Background}
\label{sec:background}
The 3D Cartesian grid class directly stores and interpolates $\alpha$, $\beta^i$, $v^i$, the six independent components of $\gamma_{ij}$, and the 40 independent components of the connection coefficients $\Gamma^\alpha_{\mu\nu}$ in the same Cartesian coordinates described above. The connection coefficients are computed using $\Gamma^\alpha_{\mu\nu} = \frac{1}{2}g^{\alpha\eta}\left(g_{\mu\eta,\nu}+g_{\eta\nu,\mu}-g_{\mu\nu,\eta}\right)$, where spatial derivatives are computed via finite differencing with nearest neighbors in each direction and interpolated and time derivatives are assumed to be zero. When constructing a comoving orthonormal tetrad, the three spacelike trial vectors are chosen to be $e_{(z,\mathrm{trial})}^\alpha=\{0,0,0,1\}$,$e_{(y,\mathrm{trial})}^\alpha=\{0,0,1,0\}$, and $e_{(x,\mathrm{trial})}^\alpha=\{0,1,0,0\}$.

\subsection{Interpolation}
\label{sec:interpolation}

We perform multi-dimensional interpolation for the values and derivatives of metric quantities and neutrino-fluid interaction rates at a neutrino's position $x^\mu$ and momentum $k^\mu$. For a position in a general n-dimensonal grid, there are $2^n$ grid points that define the hyper-cube enclosing the position. We denote the left and right coordinates of those points along each coordinate direction $k$ by $x_0^k$ and $x_1^k$, respectively. We denote the value of the function at the corners $f_{i_1 i_2 \ldots i_n}$, where each $i_k$ can be either 0 (left) or 1 (right). The value of the function at $x$ linearly interpolated in each dimension is then a sum over the values of the function values at the corners multiplied by appropriate weights:
\begin{equation}
f(\mathbf{x}) = \sum_{i_1=0}^1 \sum_{i_2=0}^1 \ldots \sum_{i_n=0}^1 W_{i_1 i_2 \ldots i_n} f_{i_1 i_2 \ldots i_n}\,\,,
\end{equation}
where the weights are
\begin{equation}
\begin{aligned}
W_{i_1 i_2 \ldots i_n} &= \frac{1}{V}\prod_k \delta x^k(i_k) \,\,,\\
\delta x^k(i_k) &= \begin{cases}
x_1^k - x^k & i_k=0\\
x^k - x_0^k & i_k=1
\end{cases}\,\,,\\
V &= \prod_k (x_1^k - x_0^k)\,\,.
\end{aligned}
\end{equation}
Similarly, the derivative of the function along direction $d$ is
\begin{equation}
\partial_d f(\mathbf{x}) = \sum_{i_1=0}^1 \sum_{i_2=0}^1 \ldots \sum_{i_n=0}^1 (S_d)_{i_1 i_2 \ldots i_n} f_{i_1 i_2 \ldots i_n}\,\,,
\end{equation}
where the weights are
\begin{equation}
(S_d)_{i_1 i_2 \ldots i_n} = \frac{2i_d-1}{V} \prod_{k \neq d} \delta x^k(i_k)\,\,.
\end{equation}
The weights can be computed once for each position and used to interpolate all quantities.

As a side note, we also explored discrete discontinuous linear interpolation of variables. In this method, the values and derivatives in each direction at the cell center are stored. This has the advantage that interpolation is much faster, but also requires more storage. Interpolating in N dimensions is simply $f(\mathbf{x}) = f(\mathbf{x}_0) + \sum_{i} (x^i-x^i_0)\partial_i f(\mathbf{x}_0)$, where $\mathbf{x}_0$ is the position of the grid point nearest to $\mathbf{x}$, and $f(\mathbf{x}_0)$ and $\partial_i f(\mathbf{x}_0)$ are stored. However, we found that the discontinuities in metric quantities across cell boundaries were problematic for the integration of the neutrino momenta. In general, once the neutrino moves across a cell boundary, the jump in the metric causes the stored neutrino four-momentum to no longer be null. We tried several methods of null-normalizing the momenta when neutrinos that cross the boundaries and special integration steps for neutrinos that cross cell boundaries, but the induced errors always led to unrealistic neutrino momenta in neutrino packets that cross many boundaries like in scattering-dominated regions.

\subsection{Emission}
\label{sec:emission}
A specified number of neutrino packets $n_{\mathrm{emit},ig}$ is emitted from each grid cell labeled by index $i$, in each energy bin labeled by index $g$, and for each species. Each of these packets is given uniform random coordinates within the grid zone (uniform values of $r^2$, $\cos\theta$, and $\phi$ for 1D spherical coordinates) and a random comoving-frame frequency uniform in $\nu^3$ within the energy bin $g$. The metric and fluid quantities are interpolated to the packet's position and momentum, the local orthonormal tetrad is constructed, and an isotropic random direction is given to the packet in the tetrad frame. The packet weight is then set to 
\begin{equation}
\begin{aligned}
N_0 = & \frac{1}{n_{\mathrm{emit},ig}}c^2 B_{s} \kappa_{{\mathrm{abs}},s}4\pi \Delta\left(\frac{\nu^3}{3}\right)_g \mathcal{V}_i \,\,,
\end{aligned}
\end{equation}
where $s$ is the species index and $i$ is the spatial grid cell index. The effective zone four-volume (i.e., that using interpolated metric quantities) is $\mathcal{V}_i = V_{\mathrm{coord},i} \sqrt{\det(\mathbf{\gamma})} (-n^\alpha u_\alpha) \Delta t$, where $V_{\mathrm{coord},i}$ is the coordinate volume of the grid cell $i$ {and $n^\alpha$ is the unit vector normal to the time slice (equivalently, the four-velocity of an Eulerian observer)}. We multiply this by the square root of the determinant of the three-metric $\mathbf{\gamma}$, the Lorentz factor $-n^\alpha u_\alpha$, and an arbitrary coordinate time interval $\Delta t=1\,\mathrm{s}$ to get the comoving four-volume of the grid cell. The $\Delta t$ is arbitrary because all quantities are divided by $\Delta t$ to yield rates, so it always cancels out. $\kappa_{\mathrm{abs},s}(\nu,\mathbf{x})$ is the absorption opacity of species $s$. In order to maintain consistency, instead of interpolating the emissivity, we interpolate the fluid temperature $T$ and electron neutrino equilibrium chemical potential $\mu_{\nu_e}=\mu_e + \mu_p - \mu_n$ given by the equation of state. We then compute the emissivity as the product of the absorption opacity with the blackbody function
\begin{equation}
    B_s(\nu,T,\mu_s) = \frac{1}{1+\exp[(h\nu-\mu_s)/k_B T]}\,\,.
\end{equation}
The chemical potentials of the different species are determined by $\mu_{\bar{\nu}_e}=-\mu_{\nu_e}$ and $\mu_{\nu_x}=0$.

We record the contribution to the volume-specific four-force exerted by the neutrino radiation on the fluid via emission $\mathcal{F}_{\mathrm{emit},i}^\alpha$ and rate of change of lepton number $\mathcal{L}_{\mathrm{emit},i}$ in spatial grid zone $i$ in the comoving tetrad frame as
\begin{equation}
\label{eq:fourforce_emission}
\begin{aligned}
    \delta \mathcal{F}_{\mathrm{emit},i}^\alpha &= -\frac{1}{\mathcal{V}_i} N_0 k^\alpha_{\mathrm{tet},q}\,\,,\\
    \delta \mathcal{L}_{\mathrm{emit},i} &= -\frac{1}{\mathcal{V}_i} N_0 l_s\,\,,
\end{aligned}
\end{equation}
where $l_s$ is the lepton number of species $s$ (1 for $\nu_e$, -1 for $\bar{\nu}_e$, and 0 for $\nu_x$).


\subsection{Standard Transport}
\label{sec:propagation}
Here we describe the standard method, used where the transport is not scattering-dominated, but will describe the scattering-dominated method in Section~\ref{sec:randomwalk}.

Each neutrino packet will take a series of small steps of length determined by the neutrino-fluid interaction rates and the distance from fluid grid cell boundaries. We express the path length of the neutrino packet in terms of a distance in the comoving tetrad $ds_\mathrm{move}$, which is a proxy for the interval in the affine parameter along the trajectory $d\lambda = ds_\mathrm{move}/k_\mathrm{tet}^t$. Each time the particle moves, the distance it moves is the smaller of the grid distance and an interaction distance. That is,
\begin{equation}
\label{eq:ds_move}
    ds_\mathrm{move} = \min(ds_\mathrm{grid}, ds_\mathrm{interact})\,\,.
\end{equation}
The tetrad-frame distance to the next scattering event is randomly sampled as
\begin{equation}
    ds_\mathrm{interact}=-(\ln U)/\kappa_\mathrm{scat}\,\,,
\end{equation}
where U is a uniform random number between 0 and 1.

In order to limit the sizes of the step to not be too large (to appropriately sample each grid cell) and not too small (to prevent spending computer time on particles close to the boundary), we set the grid distance to
\begin{equation}
\label{eq:ds_grid}
ds_\mathrm{grid} = \min(\max(ds_\mathrm{boundary},ds_{\min}), ds_{\max})\,\,.
\end{equation}
We estimate for the comoving tetrad-frame distance to the next grid cell boundary $ds_\mathrm{boundary}$ as 
\begin{equation}
ds_\mathrm{boundary} \approx k_\mathrm{tet}^t \min_i\left[ (x^i - x^i_\pm) / (g_{\mu\nu}e_{(i)}^\mu k^\nu)\right]\,\,,
\end{equation}
where $i$ ranges from 1 to the number of dimensions in the background fluid profile. $x^i_\pm$ is the grid cell boundary coordinate to the right/left of $x^i$ if $g_{\mu\nu}e_{(i)}^\mu k^\nu$ is larger/smaller than 0, respectively. In spherical symmetry, for example, this becomes $ds_\mathrm{boundary} = k_\mathrm{tet}^t (r-r_\pm) / (g_{\mu\nu}e_{(r)}^\mu k^\nu)$.  We use $ds_{\min}=0.05 ds_\mathrm{zone}$ and $ds_{\max}=0.5 ds_\mathrm{zone}$, where
\begin{equation}
    ds_\mathrm{zone} = k_\mathrm{tet}^t \min_i\left[ (x^i_+ - x^i_-) / (g_{\mu\nu}e_{(i)}^\mu k^\nu)\right]\,\,.
\end{equation}

We integrate the particle position and momentum using a kick-drift-kick method. That is, to find the particle position and momentum at step $q+1$ based on the position and momentum at step q, we use
\begin{equation}
\begin{aligned}
    k^\alpha_{q+1/2} &= k^\alpha_q - \frac{d\lambda}{2}\Gamma^\alpha_{\mu\nu}(\mathbf{x}_q)k^\mu_q k^\nu_q \,\,, \\
    x^\alpha_{q+1} &= x^\alpha_q + k^\alpha_{q+1/2}\,\,,\\
    k^\alpha_{q+1} &= k^\alpha_{q+1/2} - \frac{d\lambda}{2}\Gamma^\alpha_{\mu\nu}(\mathbf{x}_{q+1})k^\mu_{q+1/2} k^\nu_{q+1/2}\,\,.
    \end{aligned}
\end{equation}
We then renormalize $k^\alpha$ by scaling each spatial component of $k^\alpha$ by the same factor to ensure that $k^\alpha k_\alpha=0$. {In principle, not all four components of the four-momentum are independent, constrained by the
 requirement that the vector remain null. One could, for example, just integrate the spatial components and set the time component to ensure the vector is null. However, this can cause the truncation error to preferentially go to the time component.} We scale the spatial components, since we find that the errors introduced by scaling the time component can lead some neutrino packets to have unrealistically large momenta. { In addition, for a static spacetime one can in principle leverage the fact that $k_t$ must remain constant, resulting in only two independent quantities. However, \cite{Dolence2009} discuss that straightforward geodesic integration tends to be simpler to implement and modify and also faster.}

In the 3D Cartesian calculations in Section~\ref{sec:results}, there is a reflection symmetry boundary on the $z=0$ plane. The other boundaries are outflow boundaries. If the packet passes through the reflection boundary, the $z$-component of $k_\alpha$ and $x^\alpha$ are negated. The neutrino packet is immediately destroyed if $g_{tt}(\mathbf{x}_\mathrm{new})\geq 0$ (i.e., the packet passed a coordinate singularity). Finally, the packet is destroyed when it passes through an outflow boundary.

\subsubsection{Absorption}
\label{sec:absorption}
Following propagation, the packet weight is then reduced to $N_{q+1} = N_q e^{-\eta}$ to account for absorption. We calculate the absorption optical depth as $\eta = \kappa_{\mathrm{abs},q} ds_\mathrm{move}$. We record the contribution to the volume-specific four-force exerted by the neutrino radiation on the fluid via absorption $\mathcal{F}_{\mathrm{abs},i}^\alpha$ and rate of change of lepton number $\mathcal{L}_{\mathrm{abs},i}$ in spatial grid zone $i$ in the comoving tetrad frame as
\begin{equation}
\label{eq:absorption}
\begin{aligned}
    \delta \mathcal{F}_{\mathrm{abs},i}^\alpha &= \frac{1}{\mathcal{V}_i} (N_q-N_{q+1}) k^\alpha_{\mathrm{tet},q}\,\,,\\
    \delta \mathcal{L}_{\mathrm{abs},i} &= \frac{1}{\mathcal{V}_i} (N_q-N_{q+1}) l_s\,\,,
\end{aligned}
\end{equation}
where $l_s$ is the lepton number of species $s$ (1 for $\nu_e$, -1 for $\bar{\nu}_e$, and 0 for $\nu_x$).

Since absorption continuously decreases the packet weight, we roulette the packet if its weight becomes too low. That is, if a packet's weight decreases below $10^{-3}N_0$, a uniform random number $U$ between 0 and 1 is sampled. If $U<0.5$ the packet is destroyed and if $U>0.5$ the packet weight doubles. This preserves all statistical averages and prevents unimportant packets from using computer time.

\subsubsection{Neutrino Radiation Field}
\label{sec:sedonu_radiation_field}
As the packet moves, it contributes to the local radiation field. We account for this in the comoving tetrad frame where the neutrino distribution is binned into the same comoving-frame frequency bins used to discretize and store the interaction rates.
The angular structure of the distribution function $f_\epsilon$, where $\epsilon=-k^\alpha u_\alpha$ is the neutrino comoving-frame energy, is decomposed into comoving tetrad-frame moments as
\begin{equation}
\label{eq:moments}
\begin{aligned}
	E &= \frac{1}{(hc)^3} \int \frac{d\epsilon^3}{3} \epsilon\int d\Omega f_\epsilon\,\,, \\
    F^a &= \frac{1}{(hc)^3} \int \frac{d\epsilon^3}{3} \epsilon\int d\Omega f_\epsilon l^a  \,\,, \\
    P^{ab} &= \frac{1}{(hc)^3} \int \frac{d\epsilon^3}{3} \epsilon\int d\Omega f_\epsilon l^a l^b \,\,, \\
    L^{abc} &= \frac{1}{(hc)^3} \int \frac{d\epsilon^3}{3} \epsilon\int d\Omega f_\epsilon l^a l^b l^c  \,\,, \\
    ...
\end{aligned}
\end{equation}
where $l^i$ are the spatial basis vectors in the tetrad coordinates {and the differential $d\epsilon^3/3$ is equivalent to $\epsilon^2 d\epsilon$ used to integrate spherical volumes}. Of course, by construction, the basis vectors in these coordinates are simply $\{1,0,0\}$, $\{0,1,0\}$, and $\{0,0,1\}$. During a propagation step $q$, the contribution to the radiation field moments to the currently occupied spatial grid zone $i$ is then given by
\begin{equation}
\begin{aligned}
    \langle N \rangle &= N_0\int_0^\eta e^{-\eta'}d\eta' \\
    &\approx \begin{cases}
    (N_{q+1}+N_q)/2 & \eta << 1 \\
    (N_{q+1}-N_q)/\eta & \mathrm{otherwise}
    \end{cases}\\
	\delta E_{i,q} &\approx  \frac{\langle N\rangle k^t_{\mathrm{tet},q} ds_\mathrm{move}}{c\mathcal{V}_i} \\
	\delta F_{i,q}^a &= \delta E_{i,q} \frac{k^a_{\mathrm{tet},q}}{k^t_{\mathrm{tet},q}}\,\,,\\
	\delta P_{i,q}^{ab} &= \delta E_{i,q} \frac{k^a_{\mathrm{tet},q}}{k^t_{\mathrm{tet},q}}\frac{k^b_{\mathrm{tet},q}}{k^t_{\mathrm{tet},q}}\,\,,\\
	\delta L_{i,q}^{abc} &= \delta E_{i,q} \frac{k^a_{\mathrm{tet},q}}{k^t_{\mathrm{tet},q}}\frac{k^b_{\mathrm{tet},q}}{k^t_{\mathrm{tet},q}}\frac{k^c_{\mathrm{tet},q}}{k^t_{\mathrm{tet},q}}\,\,.\\
\end{aligned}
\end{equation}
The neutrino momentum is evaluated at the beginning of the step rather than the end because in the event of near complete absorption immediately following emission, all of the energy is emitted and deposited at the same location, helping avoid some noise from stiff source terms.

\subsubsection{Scattering}
\label{sec:scattering}
If the shorter of the distances in Equation~\ref{eq:ds_move} was the interaction distance, the particle will undergo an elastic scattering event. The particle is given an isotropic random direction in the comoving tetrad frame, keeping the same energy in that frame, and the new lab-frame momentum is determined by Lorentz-transforming to the lab frame (Equation~\ref{eq:tet_transform}). The resulting four-force on the fluid due to scattering is then accumulated as
\begin{equation}
    \label{eq:scattering}
    \delta \mathcal{F}_{\mathrm{abs},i}^\alpha = \frac{1}{\mathcal{V}_i} (N_{q+1} k^\alpha_{\mathrm{tet},q+1}-N'_{q+1}{k'}^\alpha_{\mathrm{tet},q+1}) \\
\end{equation}
where the prime and unprimed variables refes to the state before and after the scattering event, respectively.

As a side note, we have also experimented with biasing procedures to change the probability of scattering (e.g., to help escape scattering-dominated regions), but this inevitably caused the weights of some particles to randomly undergo scattering events several times in such a way that their weights increased to excessively large values, increasing overall variance in the solution.

\subsection{Random Walk Monte Carlo}
\label{sec:randomwalk}
It is immensely computationally inefficient to simulate a particle directly if it is in a location with a large scattering opacity, since the distance moved between scatters is extremely small. Following \cite{Fleck1984, Richers2017a,Foucart2018}, we approximate a large number of scatters with a single large step in a random direction and an appropriate random sampling of the time it took to get there. Although the salient features of this method were already outlined in \cite{Richers2017a,Foucart2018}, there are some subtle differences, and so for completeness we will describe the full method.

It was shown in \cite{Richers2017a} that if we define a sphere of radius $R$ centered around a particle's position in a homogeneous isotropic medium, the probability of escaping the sphere before time $\tau$ is
\begin{equation}
  \begin{aligned}
    P_\mathrm{escape}(R,\tau) &= 1-2\sum_{n=1}^\infty (-1)^{n-1}\\
    &\times \exp\left[-(n\pi)^2\zeta\right]
  \end{aligned}
\end{equation}
Here, $\zeta=D\tau/R^2$ and $D=c/3\kappa_\mathrm{scat}$ is the diffusion constant. We tabulate this function using 100 evenly spaced points in $\zeta$ up to $\zeta_\mathrm{max}=2$. To sample the time required to escape $\tau_\mathrm{esc,tet}$ the sphere in the comoving tetrad frame, we sample a uniform random number between 0 and 1 to substitute in for $P_\mathrm{escape}$ and invert the function numerically via linear interpolation.

Following \cite{Foucart2018}, we assume that the particle spends a period of time $\tau_\mathrm{trap} = \tau_\mathrm{esc}-R/c$ trapped at the center of the sphere and spends the remaining time $\tau_\mathrm{free}=R/c$ free-streaming to the edge of the sphere. This is not formally correct and will not account for adiabatic losses or other effects that depends on derivatives of the fluid quantities, but relativistic effects (e.g., redshift) will be approximately correct. 

We must first select an appropriate random walk sphere size $R$ before actually performing the random walk. We do this by ensuring that the coordinate-frame displacement is at most approximately the distance to the cell boundaries in each of several directions. The coordinate displacement from the random walk can be estimated as
\begin{equation}
    \Delta x^\alpha \approx \left.\frac{dx^\alpha}{d\tau}\right|_\mathrm{free} \left(\frac{R}{c}\right)+\left.\frac{dx^\alpha}{d\tau}\right|_\mathrm{trap}\left(\tau_\mathrm{esc}-\frac{R}{c}\right)
    \label{eq:dx_randomwalk_approx}
\end{equation}
where $dx^\alpha/d\tau|_\mathrm{free}=k^\alpha_{(a,\pm)}/k_{(a,\pm),\mathrm{tet}}^t$ and $dx^i/d\tau|_\mathrm{trap}=u^i$. In order to make a general algorithm that accounts for multiple possible displacement directions and the possibility of large $\zeta$, we solve Equation~\ref{eq:dx_randomwalk_approx} for $R$ with $\tau_\mathrm{esc}=R^2 \zeta_\mathrm{max}/D$, a test $k^\alpha_{(a,\pm)}$ in each coordinate direction $a$, and $\Delta x^a=x^a_\pm-x^a$, where $x^a_\pm$ is the coordinate of the right(+) or left(-) cell boundary. That is, we select the spatial components of each trial $k^\alpha_{(a,\pm)}$ to be $\pm1$ in the direction of increasing(+) or decreasing(-) coordinate $a$ and choosing a time component to make the test momentum a null vector. For a 3D Cartesian grid, $k^i_{(x,+)}=\{1,0,0\}$, $k_{(x,-)}^i=\{-1,0,0\}$, etc. For a 1D spherical grid, $k_{(r,+)}^i=\{x,y,z\}$ and $k_{(r,-)}^i=\{-x,-y,-z\}$. We then select $R$ to be the smallest of all of these trials and propagate the packet in a direction uniformly randomly selected in the tetrad frame for comoving-frame distance $c\,d\tau_\mathrm{free}$ (Section~\ref{sec:propagation}). Here we denote the four-momentum following the free-streaming step $k^{\alpha\prime}_\mathrm{free}$.

We then interpolate all fluid and metric quantities at the new position and assign a new random direction at the end of the step $k^\alpha_{q+1}$ facing outward from the surface of the sphere to cause the surface of the sphere to be isotropically bright. In order to do this, we first transform $k'_\mathrm{free}$ to the tetrad frame. We then uniformily sample a new $k_{\mathrm{tet},q+1}$ and calculate the angle between the wavevectors $\cos\Theta=k_{\mathrm{tet},q+1}^\alpha k'_{\mathrm{str,tet}_\alpha}/(k_{\mathrm{tet},q+1}^t k'_{\mathrm{str,tet},t})$ (the denominator is valid because the calculation is done in the tetrad frame). If $\cos\Theta<U$ for a uniform random number $U$ between $0$ and $1$, we reject the new wavevector and sample again until we get a wavevector that is not rejected, thereby making the surface intensity of the random walk sphere proportional to $\cos\Theta$.

We must properly account for the particle contribution to the radiation field and the four-force on the fluid. We will do this separately for the two stages (trapped and free-streaming). In the trapped stage, we assume that the neutrino contributes to the radiation field isotropically in the tetrad frame. The total energy density contributed during the trapped phase is
\begin{equation}
\begin{aligned}
  \delta E_{i,q,\mathrm{trap}} &= \frac{c\tau_\mathrm{esc}-R}{c\mathcal{V}_i}\langle N\rangle k^t_{\mathrm{tet},q}\,\,,\\
  \delta F^a_{i,q,\mathrm{trap}} &= 0\,\,,\\
  \delta P^{aa}_{i,q,\mathrm{trap}} &= \frac{\delta E_{q,i,\mathrm{trap}}}{3} \delta^{ab}\,\,,\\
  \delta L^{abc}_{i,q,\mathrm{trap}} &= 0\,\,,\\
  \end{aligned}
\end{equation}
where $\langle N\rangle$ is computed the same way as in Equation~\ref{eq:absorption} over the trapped part of the random {w}alk step. 
The energy density contribution from the free-streaming step is accounted for in the same way as in Section~\ref{sec:propagation} over the free-streaming part of the random walk step. Each time the packet changes direction (between the trapped and free-streaming steps, and following the free-streaming step), the force exerted on the fluid is accounted for in the same way as in Section~\ref{sec:scattering}.

\subsection{Neutrino Pair Annihilation}
\label{sec:sedonu_annihilation}
The simplest reconstruction of the comoving tetrad-frame distribution function from its moments (Equation~\ref{eq:moments}) is
\begin{equation}
\label{eq:distribution_expansion}
\begin{aligned}
f_\epsilon(\theta,&\phi) = f_0 + 3 f_1^i l^i  \\
&+\frac{5}{2}\left(3 f_2^{ij}l^i l^j - f_0\right)\\
&+ \frac{7}{2}\left(5f_3^{ijk}l^i l^j l^k - 3 f_1^i l^i\right) +...\,\,,
\end{aligned}
\end{equation}
where $f_0=\int d\Omega f_\epsilon{/4\pi}$, $f_1^i=\int d\Omega f_\epsilon l^i{/4\pi}$, etc. Here $l^i$ are again the values of the components of the tetrad basis vectors in the tetrad coordinates. {This is effectively a multi-dimensional extension to an expansion in Legendre polynomials, and was derived by requiring that each term be orthogonal to each other term and that each moment can be recovered by the appropriate angular integral (the angular part of Equation~\ref{eq:moments}). This expansion does not enforce that the distribution remain between 0 and 1, so many terms are required to realistically represent sharply peaked distributions. However, we use this representation simply as a tool to be able to carry out angular integrals in what follows, so the particular angular representation is not important and the ability to recover moments from angular integrals is the key property.}

The four-force on the fluid due to neutrino pair annihilation rate is an integral over both neutrino and anti-neutrino distribution functions \cite{Bruenn1985}.
\begin{equation}
\begin{aligned}
  \mathcal{F}^\mu_{(s)} &= \int \widetilde{dp}\int \widetilde{d\bar{p}} \int d\Omega\int d\bar{\Omega} p^\mu\\
  &\left(f_\epsilon\bar{f}_\epsilon \Phi^{(a)}(\cos\theta)-(1-f_\epsilon)(1-\bar{f}_\epsilon) \Phi^{(p)}(\cos\theta)\right)\,\,,
\end{aligned}
\end{equation}
where $\widetilde{dp} = d(\epsilon^3/3)/(hc)^3$. This is most easily evaluated in the local comoving tetrad. The annihilation kernel can be decomposed into Legendre polynomials:
\begin{equation}
\Phi(\mu) = \frac{1}{2}\Phi_0 + \frac{3}{2}\Phi_1 \mu + \frac{5}{2}\Phi_2 \frac{1}{2}\left(3\mu^2-1\right)+...
\end{equation}
(units of cm$^3$/s). If we integrate the annihilation four-force {using Equation~\ref{eq:distribution_expansion}} over the directions of both the neutrino and anti-neutrino distribution functions up to second order, {after a great expansion and contraction of terms} we get
\begin{equation}
\begin{aligned}
  \mathcal{F}^t_{(s)} &= (4\pi)^2\int \widetilde{dp}\int \,\,\widetilde{d\bar{p}} \,\,\epsilon \\
  &  \times\left[\frac{1}{2}(\Phi_0^{(a)}-\Phi_0^{(p)}) f_0\bar{f}_0\right. \\
  &- \frac{1}{2}\Phi_0^{(p)} \left(1-(f_0+\bar{f}_0)\right)\\ 
  &+ \frac{3}{2}(\Phi_1^{(a)}-\Phi_1^{(p)}) f_1^b\bar{f}_1^b \\
  &+ \left.\frac{5}{4}(\Phi_2^{(a)}-\Phi_2^{(p)}) \left(3 f_2^{bc}\bar{f}_2^{bc}-f_0 \bar{f}_0\right)\right]\\
  \end{aligned}
  \end{equation}
  and
  \begin{equation}
      \begin{aligned}
  \mathcal{F}^a_{(s)} &= (4\pi)^2\int \widetilde{dp}\int \widetilde{d\bar{p}} \,\,\epsilon \\
  &\times \left[\frac{1}{2}(\Phi_0^{(a)}-\Phi_0^{(p)})  f_1^a\bar{f}_0 +\frac{1}{2}\Phi_0^{(p)} f_1^a \right.\\
  &+ \frac{3}{2}(\Phi_1^{(a)}-\Phi_1^{(p)})  f_2^{ai}\bar{f}_1^i +\frac{3}{2}\Phi_1^{(p)}\frac{1}{3}\bar{f}_1^a \\
    &\left.+ \frac{5}{4}(\Phi_2^{(a)}-\Phi_2^{(p)})\left( 3f_3^{aij}\bar{f}_2^{ij}-f_1^a\bar{f}_0\right)\right]\\
\end{aligned}
\end{equation}

Integrating over energy, these become
\begin{equation}
\begin{aligned}
  \mathcal{F}^t_{(s)} &\approx \sum_{ij} \frac{1}{\bar{\epsilon}_j}\\
  &  \times\left[\frac{1}{2}(\Phi_0^{(a)}-\Phi_0^{(p)}) E_i \bar{E}_j\right. \\
  &- \frac{1}{2}\Phi_0^{(p)} \left(\mathcal{E}_i \bar{\mathcal{E}}_j-\mathcal{E}_i\bar{E}_j-E_i\bar{\mathcal{E}}_j\right)\\ 
  &+ \frac{3}{2}(\Phi_1^{(a)}-\Phi_1^{(p)}) F^b_i \bar{F}^b_j \\
  &+ \left.\frac{5}{4}(\Phi_2^{(a)}-\Phi_2^{(p)}) \left(3 P^{bc}_i\bar{P}^{bc}_j-E_i \bar{E}_j\right)\right]\\
  \end{aligned}
  \end{equation}
and
  \begin{equation}
      \begin{aligned}
  \mathcal{F}^a_{(s)} &\approx \sum_{ij} \frac{1}{\bar{\epsilon}_j}\\
  &\times \left[\frac{1}{2}(\Phi_0^{(a)}-\Phi_0^{(p)}) F^a_i\bar{E}_j +\frac{1}{2}\Phi_0^{(p)}F_i^a \bar{\mathcal{E}}_j\right.\\
  &+ \frac{3}{2}(\Phi_1^{(a)}-\Phi_1^{(p)})  P_i^{ab}\bar{F}_j^b +\frac{3}{2}\Phi_1^{(p)}\frac{1}{3}\mathcal{E}_i\bar{F}_j^a\\
    &\left.+ \frac{5}{4}(\Phi_2^{(a)}-\Phi_2^{(p)})(3 L_i^{abc}\bar{P}_j^{bc}-F_i^a\bar{E}_j)\right]\,\,,\\
\end{aligned}
\label{eq:annihil_fourforce}
\end{equation}
where $\mathcal{E}_i=\int \widetilde{dp} \epsilon\int d\Omega \,1=\pi(\epsilon_{i+1/2}^4-\epsilon_{i-1/2}^4)/(hc)^3$ is the maximum possible neutrino energy density contribution from energy bin $i$.

We use NuLib \cite{OConnor2015} to generate neutrino pair annihilation kernels $\Phi$, but only the first two moments of the kernel are given. We can guess a second moment by requiring that the annihilation rate for comoving neutrinos is zero. That is, requiring that
\begin{equation}
\Phi(\mu=1) = \frac{1}{2}\Phi_0 + \frac{3}{2}\Phi_1 + \frac{5}{2}\Phi_2 = 0
\end{equation}
and that moments of the annihilation kernel higher than the second are zero implies that
\begin{equation}
  \Phi_2 = -\frac{1}{5}\left(\Phi_0 + 3\Phi_1\right)
  \label{eq:phiannihil2}
\end{equation}
The angular dependence of the annihilation rate in vacuum is proportional to $(1-\cos\Theta)^2$, where $\Theta$ is the angle between the directions of the annihilating neutrinos, so in vacuum this approximation becomes exact.

\section{Analytic Moment Closures}
\label{sec:closure}
The general relativistic transport equations for angular moments of the radiation field moments $\widetilde{M}$ are described in detail in \cite{Shibata2011,Cardall2013}. Ignoring details and suppressing indices, the structure of these equations follows
\begin{equation}
    \begin{aligned}
        \partial_t(\sqrt{\gamma}\widetilde{M}_{(1)}) &+ \partial_j A(g,\widetilde{M}_{(2)}) + \partial_\epsilon B(g,\widetilde{M}_{(3)},\nabla u) \\
        &= \mathcal{S}(g,\nabla g,\widetilde{M}_{(2)})\,\,.\\
    \end{aligned}
    \label{eq:moment_transport}
\end{equation}
The subscript in parentheses indicates the rank of the moment. That is $\widetilde{M}_{0}$ is the energy density, $\widetilde{M}_{(1)}$ contains the energy density and flux vector, $\widetilde{M}_{(2)}$ contains the energy density, flux vector, and pressure tensor, etc. These will be defined more carefully in the comoving orthonormal tetrad below. $A$,$B$, and $\mathcal{S}$ are functions whose details are not important for our purposes, except for the dependencies indicated in the function arguments. Most importantly, the evolution equation for the rank-1 tensor depends on the rank-2 and rank-3 tensors, which are not independently evolved and must be estimated by some other means.

The lab-frame moment tensors $\widetilde{M}$ can be constructed from from the moments in the comoving orthonormal tetrad defined in Equation~\ref{eq:moments} using tetrad basis vectors $e^{(\alpha)}_\mu$
\begin{equation}
\begin{aligned}
    \widetilde{M}_{(3)}^{\alpha\beta\gamma}&=M^{\mu\nu\eta}e_\mu^{(\alpha)}e_\nu^{(\beta)}e_\eta^{(\gamma)}\,\,,\\
    \widetilde{M}_{(2)}^{\alpha\beta}&=M^{\mu\nu}e_\mu^{(\alpha)}e_\nu^{(\beta)}\,\,,\\
    \widetilde{M}_{(1)}^{\alpha}&=M^{\mu}e_\mu^{(\alpha)}\,\,,\\
    \end{aligned}
\end{equation}
where $e^{(\alpha)}$ are the set of four tetrad basis vectors. The moments in the comoving orthonormal tetrad take the form of
\begin{equation}
\begin{aligned}
    M^{ijk} &= L^{ijk}\,\,,\\
    M^{tij}&=M^{ij}=P^{ij}\,\,,\\
    M^{tti}&=M^{ti}=M^i=F^i\,\,,\\
    M^{ttt}&=M^{tt}=M^t=E\,\,.
    \end{aligned}
\end{equation}
Note that $M^{\alpha\beta}_{\phantom{\alpha\beta}\beta}=0$. The closure to Equation~\ref{eq:moment_transport} (i.e., determining the rank-2 and rank-3 tensors) is usually implemented in the comoving orthonormal tetrad, so to have a well-defined set of evolution equations we need a prescription for the unknown comoving tetrad moments $P^{ij}$ and $L^{ijk}$ in terms of known quantities. 

All of the closures used in the literature rely on a few basic assumptions, which we will assess Section~\ref{sec:results}. The pressure tensor at each neutrino energy $\epsilon$ is assumed to take the form
\begin{equation}
\begin{aligned}
P^{ij} &= \frac{3\chi_p -1}{2}P^{ij}_\mathrm{free} + \frac{3(1-\chi_p)}{2}P^{ij}_\mathrm{diff}\,\,, \\
L^{ijk} &= \frac{3\chi_l -1}{2}L^{ijk}_\mathrm{free} + \frac{3(1-\chi_l)}{2}L^{ijk}_\mathrm{diff}\,\,, \\
\end{aligned}
\label{eq:closure_interpolation}
\end{equation}
where, under the regular assumptions that the radiation field is symmetric about the flux direction
\begin{equation}
\begin{aligned}
P^{ij}_\mathrm{free} &= E_\epsilon \frac{F_\epsilon^i F_\epsilon^j}{\mathbf{F}_\epsilon \cdot \mathbf{F}_\epsilon}\,\,, \\
P^{ij}_\mathrm{diff} &= \frac{E_\epsilon}{3}I^{ij}\,\,,\\
L_{iii,\mathrm{diff}} &= \frac{3 F_i}{5}\,\,,\\
L_{ijj,\mathrm{diff}} &= \frac{F_i}{5}\,\,,\\
L_{ijk,\mathrm{diff}} &= 0\,\,,\\
L_{ijk,\mathrm{free}} &= \frac{F_i F_j F_k}{\sqrt{F_i F^i}}\,\,.
\end{aligned}
\label{eq:PL_diff_free}
\end{equation}
Different analytical closures differ in how they interpolate between the diffusive and free-streaming limits based on the flux factor $\xi_\epsilon = \sqrt{\mathbf{F}_\epsilon \cdot \mathbf{F}_\epsilon / E_\epsilon^2}$. With these quantities defined in an orthonormal tetrad moving with the fluid, they can be transformed out using the tetrad basis vectors.

\subsection{Extending the MEFD Closure}
\label{sec:analytic_MEFD}
It is often assumed that the interpolating function between the diffusive and free-streaming limits for the third moment is the same as for the second moment. \cite{Banach2013} also used the maximum entropy condition to generate a closure for the third moment. However, the closure was expressed as a power series, so it is only accurate near the diffusion limit. In addition, \cite{Banach2017} writes limiting cases of the closure, but does not write it down in generality. In both cases, the closures for the third moment are designed for so-called nine-moment systems, in which the energy density, three fluxes, and five independent components of the pressure tensor are evolved variables (they would call our case a four-moment system, since we try to evolve the energy density and three fluxes). As such, the closure is derived using a functional for the distribution function that has more free variables, resulting in a closure that depends on all three of the energy density, flux, and pressure. However, we wish to use a closure for a two-moment system that has only two independent variables. 

To do this, we follow \cite{Cernohorsky1994} and derive an approximation to a maximum-entropy closure for the third moment based on only the energy density and flux. The MEFD closure maximizes the entropy for a{n} angular distribution with functional form
\begin{equation}
    f_\mathrm{{\epsilon}}(\mu) = \frac{N}{e^{\eta-a\mu}+k}\,\,,
    \label{eq:f_MEFD}
\end{equation}
where $\eta$ and $a$ are parameters that determine the angular distribution and $k=-1$ is used for Bose-Einstein statistics and $k=1$ is used for Fermi-Dirac statistics. The angular moment integrals are then
\begin{equation}
    \begin{aligned}
        f &= \frac{1}{e}\frac{1}{4\pi}\int_0^{2\pi} d\phi \int_{-1}^1 d\mu \mu f(\mu)\\
        p &= \frac{1}{e}\frac{1}{4\pi}\int_0^{2\pi} d\phi \int_{-1}^1 d\mu \mu^2 f(\mu)\\
        l &= \frac{1}{e}\frac{1}{4\pi}\int_0^{2\pi} d\phi \int_{-1}^1 d\mu \mu^3 f(\mu)\,\,,\\
    \end{aligned}
    \label{eq:MEFD_moment_integrals}
\end{equation}
where we are using for shorthand $e=E/E_\mathrm{max}$ (the occupation probability), $f=|F|/E$ (the flux factor, distinct from the distribution function $f_{{\epsilon}}$), $p=P_{ff}/E$, and $l=L_{fff}/E$. $N$ is a normalization factor that cancels out everywhere in this analysis. Finding a closure amounts to solving for $\eta(e,f)$ and $a(e,f)$. {In the classical limit ($e\rightarrow 0$ or k=0), these integrals are analytic, yielding $f= \coth(a)-1/a$, $p= 1-2f/a$, and $l= \left[(6+a)^2 f - 2\right]/a^2$. However, for general Fermi-Dirac radiation case we must first look at limiting cases.}

We can then evaluate the integrals under the assumption of maximum packing. That is, assuming the distribution is $f=1$ between $\mu=1$ and $\mu=\mu_0$ and $0$ outside of that range. Under these assumptions, the same moments become
\begin{equation}
    \begin{aligned}
        f_\mathrm{max}(e) &= 1-e\,\,, \\
        p_\mathrm{max}(e) &= \frac{2(1-e)(1-2e)}{3} + \frac{1}{3}\,\,, \\
        l_\mathrm{max}(e) &= (1-e)(1-2e+2e^2)\,\,.
    \end{aligned}
    \label{eq:MEFD_max}
\end{equation}

For functional form of the distribution function in Equation~\ref{eq:f_MEFD}, we can in general express the pressure and third moment in terms of the flux saturation $x=f/f_\mathrm{max}$ as
\begin{equation}
    \begin{aligned}
        p(e,x) &=\left[p_\mathrm{max}(e)-p_\mathrm{diff}(e,1)\right]\zeta_p(e,x)+p_\mathrm{diff}(e,x)\,\,,\\
        l(e,x) &= \left[l_\mathrm{max}(e)-l_\mathrm{diff}(e,1)\right] \zeta_l(e,x) + l_\mathrm{diff}(e,x)\,\,,
    \end{aligned}
    \label{eq:MEFD_closure}
\end{equation}
where the diffusive solution is $p_\mathrm{diff}(e,x)=1/3$ and $l_\mathrm{diff}(e,x)=3 x f_\mathrm{max}(e)/5$. The functions $\zeta_p(e,x)$ and $\zeta_l(e,x)$ are not representable analytically, requiring numerical root finding to get $a(e,f)$. However, we can analytically express both in the isotropic ($x\rightarrow 0$) and high-packing ($x\rightarrow 1$) limits.

Following \cite{Cernohorsky1994}, we can approximate $f(\mu,\eta,a)$ using the first two terms of a Sommerfeld expansion to get the high-packing limit. We arrive at
\begin{equation}
    \begin{aligned}
        x_{x\rightarrow1}&\approx1-\frac{A}{a^2}\,\,,\\
        \zeta_{p,x\rightarrow1} &\approx 1-\frac{3A}{a^2}\,\,,\\
        \zeta_{l,x\rightarrow1} &\approx 1-\frac{3A(1-2e)^2+3a^2 x/5}{l_\mathrm{max}(e)/f_\mathrm{max}(e)-3/5}\,\,,\\
    \end{aligned}
\end{equation}
where $A(e)=\pi^2/[12e(1-e)]$. After eliminating $a$, this becomes
\begin{equation}
    \begin{aligned}
        \zeta_{p,x\rightarrow1} &\approx 3x-2\,\,,\\
        \zeta_{l,x\rightarrow1} &\approx 6x-5\,\,.\\        
    \end{aligned}
    \label{eq:zeta_0}
\end{equation}

Again following \cite{Cernohorsky1994}, in the isotropic limit $a<<1$ and $f(\mu,\eta,a)$ can be Taylor-expanded around $a=0$ keeping only the first two terms. Including three terms gives the same result, and four or more yields intractable expressions. This leads to
\begin{equation}
    \begin{aligned}
        x_{x\rightarrow0} &\approx \frac{a}{3}\,\,,\\
        \zeta_{p,x\rightarrow0} &\approx \frac{a^2}{15}\,\,, \\
        \zeta_{l,x\rightarrow0} &\approx \frac{a/5 - 3x/5}{l_\mathrm{max}(e)/f_\mathrm{max}(e)-3/5}\,\,.
    \end{aligned}
\end{equation}
Again eliminating $a$, this becomes
\begin{equation}
    \begin{aligned}
        \zeta_{p,x\rightarrow0} &\approx \frac{3x^2}{5} \,\,,\\
        \zeta_{l,x\rightarrow0} &\approx 0\,\,.
    \end{aligned}
    \label{eq:zeta_1}
\end{equation}

\begin{figure}
    \centering
    \includegraphics[width=\linewidth]{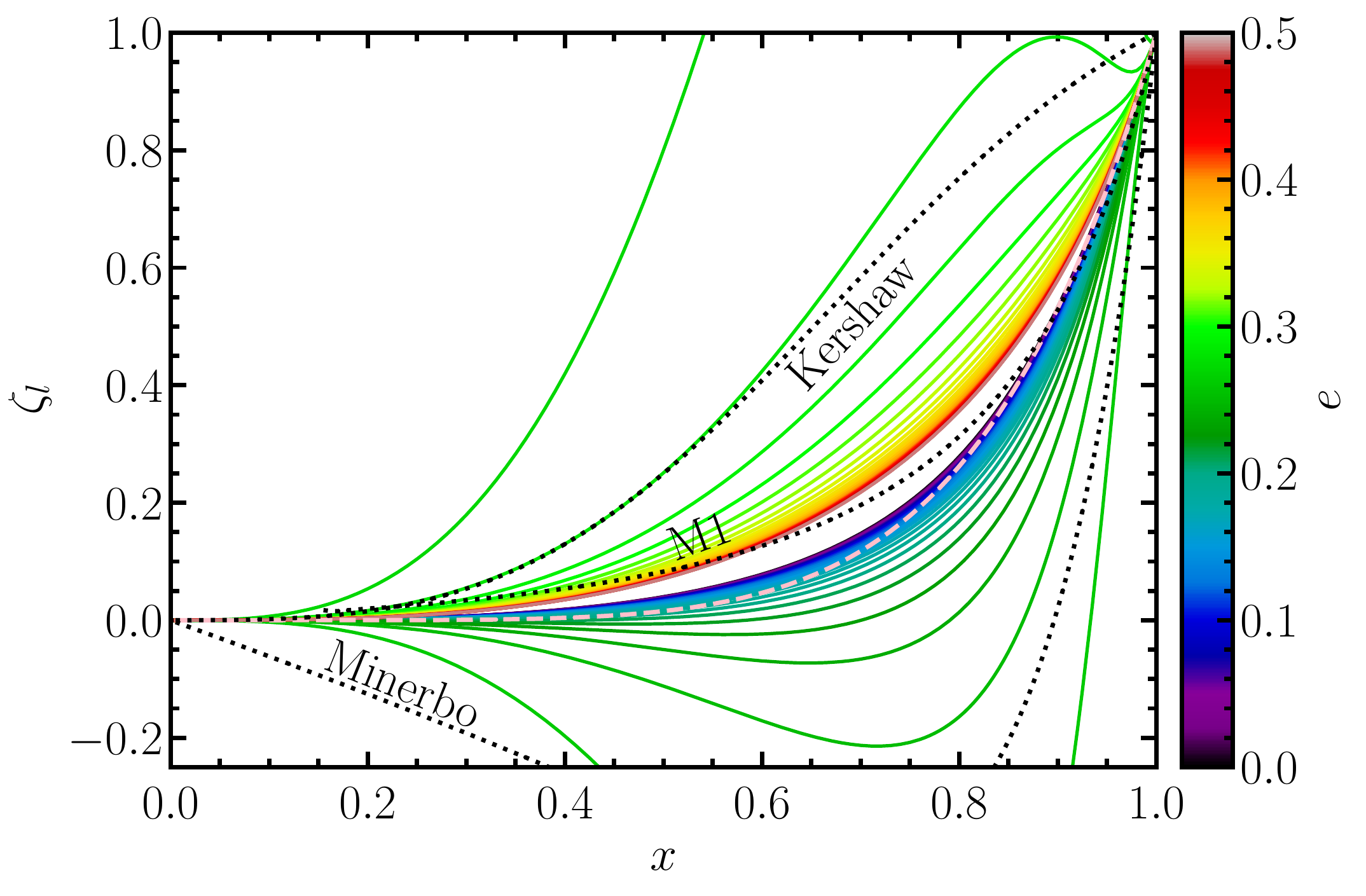}
    \caption{MEFD rank-3 radiation saturation curve $\zeta_l$ as a function of the flux saturation $x=f(x,e)/f_\mathrm{max}(e)$ (implicitly defined in Equation~\ref{eq:MEFD_closure}). Different colors correspond to different energy saturation $e=E/E_\mathrm{max}$ (effectively the direction-averaged occupation number). All of the curves have the same value and derivative in the limits of $x\rightarrow0$ and $x\rightarrow1$. The dashed white curve shows the approximant in Equation~\ref{eq:MEFD_approximate}. {The lower dotted curve shows an approximation for $\zeta_l$ as derived for the Minerbo closure (equivalent to the classical limit of the MEFD closure) by \cite{Just2015} using a different approximation for the Langevin function. The middle dotted curve shows the corresponding curve for the M1 closure as derived by \cite{Vaytet2011}. The upper dotted curve shows a suggestion for the rank-3 Kershaw closure from taking $\chi_l = \chi_p$.}}
    \label{fig:zeta_l}
\end{figure}
Both $\zeta_p$ and $\zeta_l$ must be invariant under $e\leftrightarrow (1-e)$. We see from Equations~\ref{eq:zeta_1} and \ref{eq:zeta_0} that in the isotropic and high-packing limits both are in fact independent of $e$, and \cite{Cernohorsky1994} showed that $\zeta_p(e,x)=\zeta_p(x)$ is actually completely independent of $e$. The solid curves in Figure~\ref{fig:zeta_l} shows the variation in the $\zeta_l(e,x)$ over different values in $e$.

\begin{figure}
    \centering
    \includegraphics[width=\linewidth]{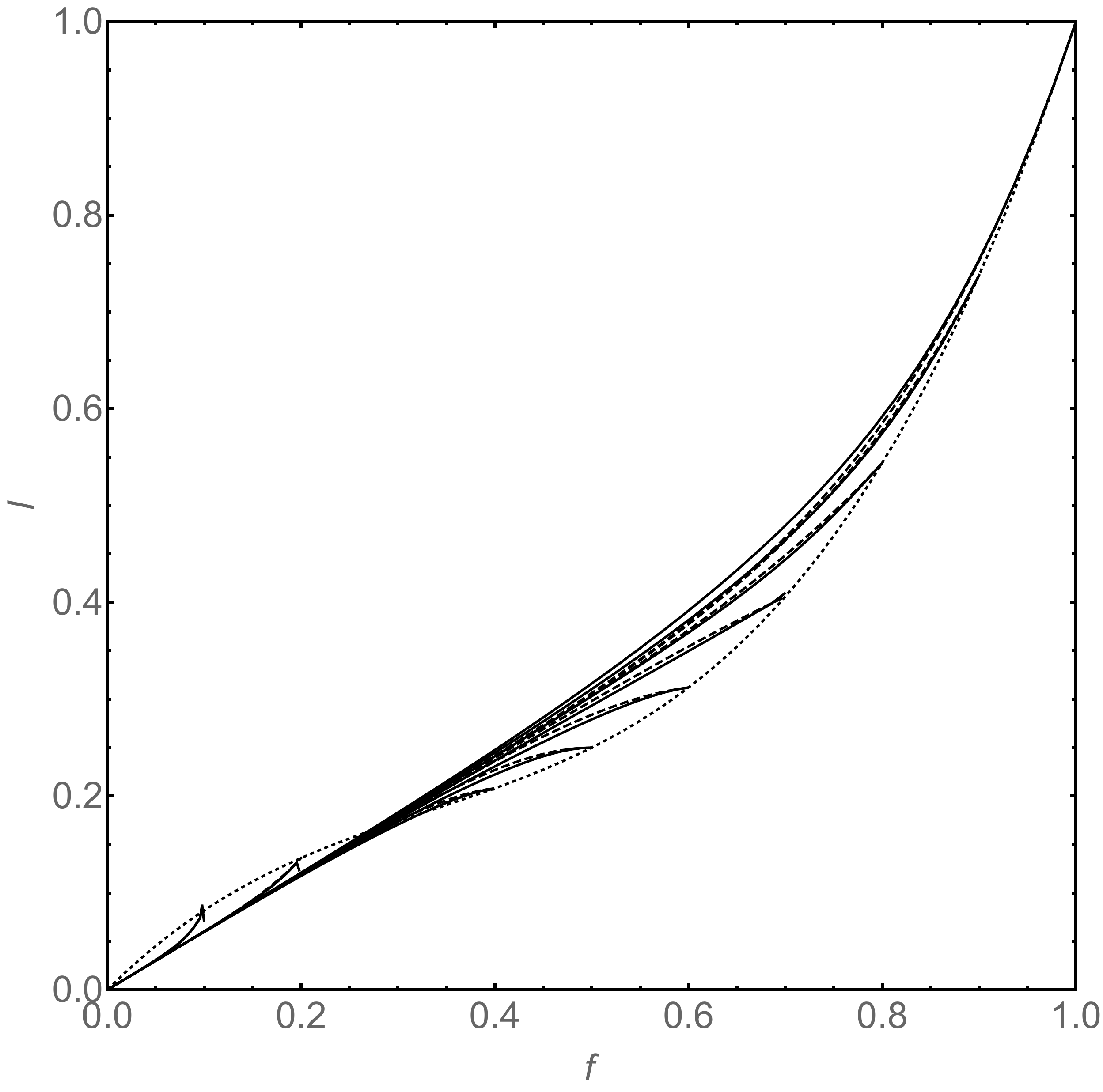}
    \caption{MEFD closure for the rank-3 radiation field tensor. $l=L_{fff}/E$ is the amount of energy in the component of the tensor aligned with the flux as a function of the flux factor $f$. Each solid lines show the results from solving for $a$ and integrating Equations~\ref{eq:MEFD_moment_integrals} for a chosen value of the energy saturation $e$. $e$ ranges from 0.1 (curve ending at $f=0.1$ to 1.0 (curve ending at $f=1.0$) in increments of 0.1. The dashed lines show the results from using Equation~\ref{eq:MEFD_closure} using Equation~\ref{eq:MEFD_approximate}. The dotted curve is the maximum packing curve (Equation~\ref{eq:MEFD_max}) that traces the endpoints of each of the curves. Note that the seemingly large deviations of the approximant in Equation~\ref{eq:MEFD_approximate} do not cause the solid and dashed lines to be far separated.}
    \label{fig:MEFD_l}
\end{figure}
$\zeta_l(e,x)$ can be approximated using the lowest-order polynomial that satisfies the values and derivatives of the functions in the high-packing and isotropic limits, along with the requirement that $0\leq \zeta(e,x) \leq 1$ (based on observation of the numerical solution).
\begin{equation}
    \begin{aligned}
        \zeta_p(e,x) &\approx x^2 (3-x+3x^2)/5\,\,, \\
        \zeta_l(e,x) &\approx x^6\,\,.
    \end{aligned}
    \label{eq:MEFD_approximate}
\end{equation}
\cite{Cernohorsky1994} showed that the approximation for $\zeta_p(x)$ is accurate everywhere within $2\%$. The dashed curve in Figure~\ref{fig:zeta_l} shows this approximation for $\zeta_l(e,x)$. While the error appears to be quite large near $e=0.3$, the prefactor in Equation~\ref{eq:MEFD_closure} is quite small. Figure~\ref{fig:MEFD_l} shows the closure curves of $l(e,f)$ (solid curves) for several values of $e$ and the corresponding approximate curve (dashed curves). The approximation produces values of $l$ that are accurate to within $3.5\%$ for any value of $f$ and $e$, and errors are largest at the $x=0,1$ and $e=0.5$ limits.

One can relate $\zeta$ back to $\chi$ based on Equation~\ref{eq:closure_interpolation} via
\begin{equation}
    \chi_p(e,x) = \frac{2}{3}\frac{p_\mathrm{max}(e)-p_\mathrm{diff}(e,1)}{p_\mathrm{free}(e,x)-p_\mathrm{diff}(e,x)}\zeta_p + \frac{1}{3}
    \label{eq:chi_from_zeta}
\end{equation}
and similarly for $\chi_l$, where $p_\mathrm{free}(e,x)=1$ and $l_\mathrm{free}(e,x)=x f_\mathrm{max}(e)$. The classical limit of the MEFD closure is obtained by taking $e\rightarrow0$ so that $f_\mathrm{max}=p_\mathrm{max}=l_\mathrm{max}=1$, leading to
\begin{equation}
    \begin{aligned}
        \chi_{p,\mathrm{classical}} &\approx \frac{2}{15}f^2\left(3-f+3f^2\right)+\frac{1}{3}\,\,,\\
        \chi_{l,\mathrm{classical}} &\approx \frac{2}{3}f^5 + \frac{1}{3} \,\,.\\
    \end{aligned}
\end{equation}
The maximum packing limit is obtained by taking $e\rightarrow 1-f$ so that $x=\zeta=1$. In this limit,
\begin{equation}
    \begin{aligned}
        \chi_{p,\mathrm{maxpack}} &\approx \frac{1}{3}(1-2f+4f^2)\,\,,\\
         \chi_{l,\mathrm{maxpack}} &\approx \frac{1}{3}(3-10f+10f^2)\,\,.
    \end{aligned}
\end{equation}

Note Equation~\ref{eq:closure_interpolation} disagrees with the choice of \cite{Shibata2011}. Our choice is made in order to always preserve the identity that $L_{ij}^j=F_i$, though an appropriate modification to the closure can ensure this indirectly. If one chooses the free-streaming limit of the third moment to be
\begin{equation}
    L_{ijk,\mathrm{free}}=\frac{J F_i F_j F_k}{(F_i F^i)^{3/2}}
    \label{eq:Lfree_BAD}
\end{equation}
then Equation~\ref{eq:chi_from_zeta} must be applied with $l_\mathrm{free}(e,x)=1$ since the interpolation is between the diffusive solution and one where all \textit{energy} (rather than flux) is moving in one direction. This leads to
\begin{equation}
    \begin{aligned}
        \chi_{l,\mathrm{classical}} &\approx \frac{2}{3}\frac{2f^6}{5-3f} + \frac{1}{3} \,\,,\\
        \chi_{l,\mathrm{maxpack}} &\approx \frac{2}{3}f\left(\frac{2-10f + 10f^2}{5-3f}\right) + \frac{1}{3}\,\,,
    \end{aligned}
    \label{eq:chil_BAD}
\end{equation}
and results in the curves in Figure~\ref{fig:MEFD_l} remaining unchanged. We will demonstrate how the choice of free streaming limit impacts other closures in Section~\ref{sec:results_L}.

{We also show $\zeta_l$ from the Minerbo closure as derived by \cite{Just2015} (converting to $\zeta_l$ from their Equations 32-33) as the lower black dotted line in Figure~\ref{fig:MEFD_l}. Their result differs considerably, although the Minerbo closure is identical to the classical limit of the MEFD closure. Although the limiting values of $\zeta_l$ at $x=\{0,1\}$ are correct, the limiting behavior differs from that derived in Equations~\ref{eq:zeta_0} and \ref{eq:zeta_1} due to a choice in how they approximate the inversion. \cite{Cernohorsky1994} choose a simple polynomial to approximate the Langevin function such that the limiting behaviors of $\zeta_p$ are correct. \cite{Just2015} use this same function to approximate the Langevin function to determine $\zeta_l$. By contrast, we follow the same process used in \cite{Cernohorsky1994} to choose a different approximation that causes the limiting behaviors of $\zeta_l$ to be correct. Both versions are valid as closures, but our classical limit exhibits smaller errors from the exact classical solution under the MEFD assumptions. For reference, we also show the equivalent curve from the M1 rank-3 closure as derived by \cite{Vaytet2011} as the upper black dotted curve in Figure~\ref{fig:MEFD_l}, converted from their Equations 12-14. There is no reason this curve should match any of the others, as it was derived under different assumptions and is just shown for comparison. We only plot values for $x>0.15$ because below this value the results of evaluating the expressions are dominated by round-off errors. Finally, the Kershaw closure \cite{Kershaw1976} can in principle be extended to the third moment in a way that obeys realizability constraints (e.g., \cite{Schneider2016}). If one assumes that $\chi_l=\chi_p$ the result fits nicely into the realizable moment space, and this can be considered the three-moment extension of the Kershaw closure. This is plotted as the upper dotted curve in Figure~\ref{fig:MEFD_l} for comparison, and also has no reason to follow any of the other curves.}

\begin{figure}
    \centering
    \includegraphics[width=\linewidth]{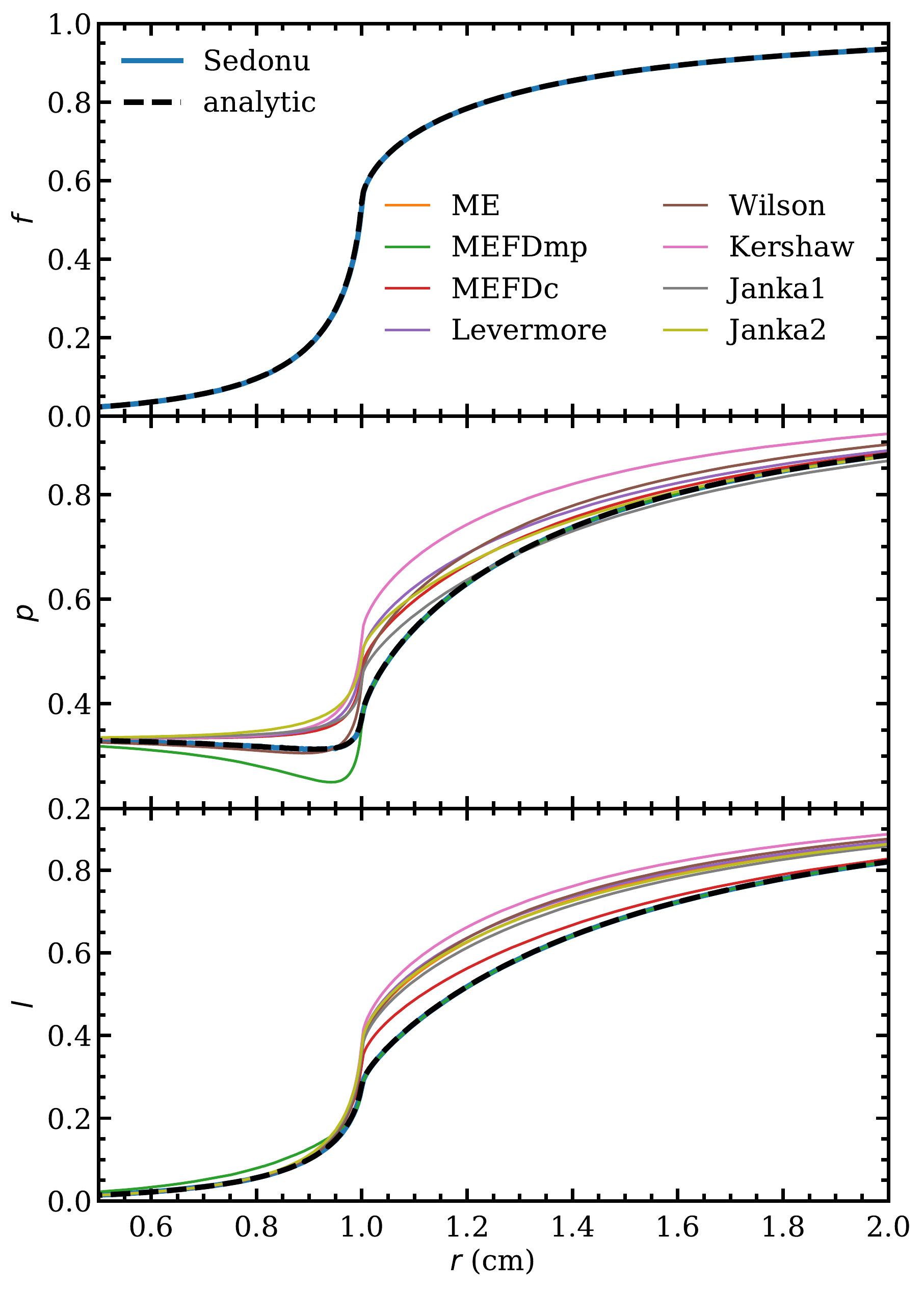}
    \caption{{Uniform sphere test. There is a constant absorption opacity of $\kappa_\mathrm{abs}=4\,\mathrm{cm}^{-1}$ below $r=1\,\mathrm{cm}$ and vacuum above. The top panel shows the resulting radial flux factor $f$, the middle panel shows the radial component of the pressure tensor $p$, and the bottom panel shows the radial component of the rank-3 tensor $l$. The analytic result (e.g., \cite{Smit1997} is shown as a black dashed curve, and the Monte Carlo results computed by Sedonu are shown with a thick blue curve. The bottom two panels also show the results from several approximate moment closures described in more detail in Section~\ref{sec:specific_closures}. No closure reproduces the physical results everywhere, though the MEFDmp closure is well suited to this test problem for $r>1\,\mathrm{cm}$.}}
    \label{fig:uniform_sphere}
\end{figure}

{As an appetizer, we present the performance of the MEFD and other popular closures (described in more detail in Section~\ref{sec:specific_closures}) in a simple test problem. As in \cite{Smit1997}, we create a homogeneous sphere with radius $R=1\,\mathrm{cm}$, a constant absorption opacity of $\kappa_\mathrm{abs}=4\,\mathrm{cm}^{-1}$, and no scattering opacity. There is an analytic solution to the radiation field (also outlined in, e.g., \cite{Smit1997}), which we plot in Figure~\ref{fig:uniform_sphere} as a black dashed curve. The thick blue curve immediately under the dashed curve is the result from Sedonu directly and shows excellent agreement. Outside of the sphere where the opacity is zero, the maximum packing limit of the MEFD closure also matches well, though it performs poorly inside the sphere. The opposite is true of most of the other closures, and already it is apparent that none of these closures performs well everywhere, as concluded by \cite{Smit2000,Murchikova2017}. The goal of this paper is to perform a similar assessment of these closures, for up to the rank-3 moments and in the multidimensional and relativistic environment of a neutron star merger.}

\subsection{Tensor Invariants}
\label{sec:tensor_invariants}
The pressure tensor in the comoving orthonormal tetrad is diagonalizable, meaning that with the proper rotation of coordinates one can express the pressure tensor with only three diagonal elements. These correspond to the neutrino pressure in three directions. 
\begin{equation}
    P^{ij} = \begin{bmatrix}
    P^{xx} & P^{xy} & P^{xz} \\
    P^{yx} & P^{yy} & P^{yz} \\
    P^{zx} & P^{zy} & P^{zz} \\
    \end{bmatrix} = R \begin{bmatrix}
    \lambda_0 & 0 & 0 \\
    0 & \lambda_1 & 0 \\
    0 & 0 & \lambda_2
    \end{bmatrix} R^T\,\,.
\end{equation}
Following \cite{Kindlmann2006a}, the eigenvalues can be expressed as 
\begin{equation}
    0=|\lambda I^{ij} - P^{ij}| = \lambda^3 - J_1 \lambda^2 + J{_2} \lambda - J{_3}\,\,,
\end{equation}
where
\begin{equation}
    \begin{aligned}
        J_1 &= \mathrm{Tr}(P^{ij})=E \,\,,\\
        J_2 &= \frac{1}{2}[\mathrm{Tr}(P^{ij})^2-\mathrm{Tr}(P^{ij}P^{jk})]\,\,,\\
        J_3 &= |P^{ij}|\,\,.
    \end{aligned}
\end{equation}
$\lambda$, $J_1$, $J_2$, and $J_3$ are all invariant under rotation. Furthermore, we can write the eigenvalues as a function of the $J$ invariants as
\begin{equation}
    \begin{aligned}
        \lambda_k &= \frac{1}{3}J_1 + 2\sqrt{Q} \cos\left(\theta + \frac{2\pi}{3}k\right) \,\,,\\
    \end{aligned}
\end{equation}
where $Q = (J_1^2-3J_2)/9$, $\theta = \left[\cos^{-1}\left(R Q^{-3/2}\right)\right]/3$, and $R = \left(23 J_3-9J_1 J_2+2J_1^3\right)/54$. Thus, the three eigenvalues can be visualized as projections onto the x-axis of three equally-spaced points on a circle centered at (E/3,0). The magnitude of the differences between the eigenvalues is determined by $Q$ and the configuration of the eigenvalues is determined by $\theta$.

In our analysis, we will refer to the dimensionless quantities
\begin{equation}
\begin{aligned}
    \mathrm{anisotropy} &= \frac{3\sqrt{Q}}{J_1} \in [0,1]\,\,,\\
    \mathrm{oblateness} &= \frac{3\theta}{\pi} \in [0,1]\,\,.
    \end{aligned}
    \label{eq:anisotropy_oblateness}
\end{equation}
The pressure tensor can be visualized as a triaxial ellipsoid, where the size of each axis represents the size of an eigenvalue, i.e., the magnitude of the pressure in the direction of the corresponding eigenvector. An anisotropy of 0 indicates that the pressure in all directions is equal, while an anisotropy of 1 indicates that the pressure in all but one direction is zero. An oblateness of 0 means the ellipsoid is prolate, i.e., that two eigenvalues are equal and one is larger. An oblateness of 1 means the ellipsoid is oblate, i.e., that two eigenvalues are equal and one is smaller. An oblatness between 0 and 1 means that no two of the eigenvalues are equal. These rotation-independent quantities are useful for understanding the limitations imposed by moment closures.

Similarly, there are 11 invariants for the rank-3 tensor $L^{ijk}$ \cite{Ahmad2011}. However, several of these invariants are degenerate or represent the know relationship between the trace of $L$ and the flux. After ignoring all of the known or degenerate invariants, only one remains, which we call $L_4$ (using the subscript from \cite{Ahmad2011}). 
\begin{equation}
    L_4 = L^{ijk}L^{ijk}\,\,.
    \label{eq:L4}
\end{equation}
We will use this as a scalar quantity representing $L$ so we can compute differences between closures without needing to refer to particular directions.

\section{Results}
\label{sec:results}
We perform time-independent Monte Carlo neutrino radiation transport on a simulation snapshot from the LS220\_M135135\_M0\_L25 simulation of \cite{Radice2018}. The snapshot is at $t=31.3\,\mathrm{ms}$ after merger and is after the formation of a black hole. The neutrino interaction rates and comoving-frame radiation field are binned into 24 energy bins spaced logarithmically from 1 MeV (bin width of 2 MeV) to 270 MeV (bin width of 37 MeV). We performed the transport on a single refinement level with Cartesian grid spacing of $0.74\,\mathrm{km}$, a domain of $-66-66\,\mathrm{km}$ in each coordinate direction, and a reflecting boundary condition at $z=0$. This refinement level was chosen to be the smallest level that contained the regions of the accretion disk where transport is relevant.

\begin{figure*}
    \centering
    \includegraphics[width=\linewidth]{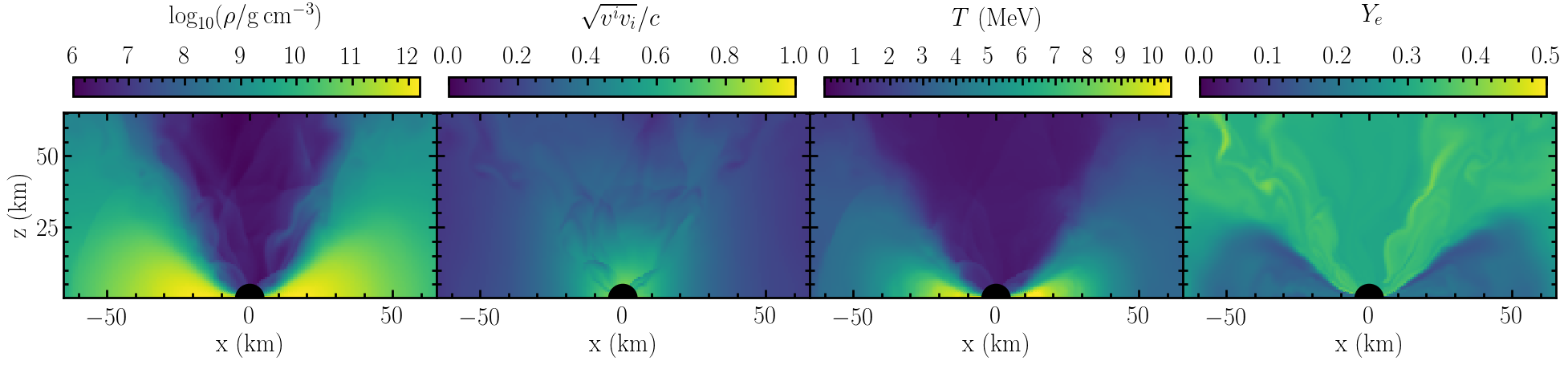}
\caption{Background fluid profile from the LS220\_M135135\_M0\_L25 simulation of \cite{Radice2018} on top of which we calculate steady-state neutrino radiation fields using Sedonu. \textit{First panel:} baryon rest density. \textit{Sedond panel:} magnitude of the three-velocity. \textit{Third panel:} fluid rest temperature. \textit{Fourth panel:} electron fraction.}
    \label{fig:profile}
\end{figure*}
A $x-z$ slice of the fluid data is shown in Figure~\ref{fig:profile}. The complicated matter, velocity, and spacetime structure pose a significant challenge for radiation transport algorithms. There is a dense emitting disk and a sparse polar region. Though the second panel in Figure~\ref{fig:profile} only shows the magnitude of the three-velocity, the large velocity in the disk is in the azimuthal direction, while the velocity in the polar region is in the positive $z$-direction. There is also a large $x$-velocity in the boundary between the two within $10-15\,\mathrm{km}$ from the black hole. The large temperature in the inner regions of the disk (third panel) indicate where most of the neutrinos are being produced, and the disk still has a low $Y_e$ (fourth panel) as antineutrino losses continue to drive up the $Y_e$.

\begin{figure}
    \centering
    \includegraphics[width=\linewidth]{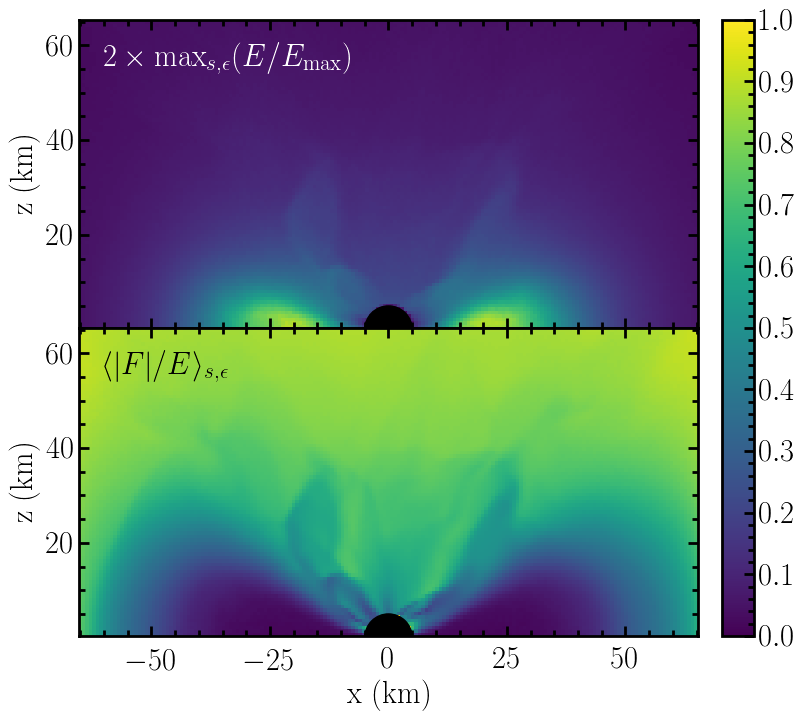}
    \caption{\textit{Top panel:} neutrino occupation number, maximized over neutrino species and neutrino energy. \textit{Bottom panel:} comoving-frame neutrino flux factor, averaged over species and neutrino energy, weighted by the energy density in the corresponding species-energy bin. Neutrinos are mildly degenerate and trapped in the disk and free-streaming in the polar regions. The goal of an analytic closure is to use just this information to predict all higher-rank moments.}
    \label{fig:e_fluxfac}
\end{figure}
We simulate $2\times10^9$ neutrino packets to generate a steady-state radiation field according to Section~\ref{sec:SedonuGR}. As in the original dynamical calculation, we use the LS220 equation of state \cite{Lattimer1991} and use NuLib \cite{OConnor2015} to calculate neutrino absorption and elastic scattering rates. The resulting comoving-frame maximum occupation number (top panel) and energy-density-averaged flux factor (bottom panel) computed by {\tt SedonuGR} are shown in Figure~\ref{fig:e_fluxfac}. The maximum occupation number of a given energy bin is computed by dividing the energy density in an energy bin by the energy density that would be present in that bin if the occupation number were 1. Though neutrinos can become very degenerate in a proto- or hypermassive neutron star, in this disk they are only mildly degenerate. The flux factor plot shows that the disk is on average optically deep (resulting in a flux factor close to 0) and the polar regions are optically thin (resulting in a flux factor approaching unity). Even with only two moments, it is apparent that there is interesting structure in the radiation field, especially in the interface between the disk and polar regions. This will prove to be a very difficult region for analytic closures.

The simulation in \cite{Radice2018} was performed with the $M0$ scheme which combines a leakage method for radiation losses with a diffusion method for neutrino reheating. The goal of the rest of this section is not to analyze the differences between the two methods (as in, e.g., \cite{Richers2015,Foucart2018a}, but to quantify the errors induced by the assumptions that go into closure relations. To do this we will compare the rank-2 and rank-3 moments with the energy density and flux, all computed in the same calculation by {\tt SedonuGR}.

\subsection{Assessing Closure Assumptions}
\label{sec:assessing_assumptions}
The analytic closure method described in Section~\ref{sec:closure} attempts to capture the dominant structure of the radiation field present in the rank-2 and rank-3 moments. In this section, we use {\tt SedonuGR} to assess the ability of such a closure to represent the real second and third angular moments of the radiation field. While other authors have compared the results of simulations performed using Monte Carlo and moment methods (e.g. \cite{Foucart2018a}), here we focus instead on how well the Monte Carlo radiation field respects the fundamental assumptions that go into forming such a closure. These assumptions are (1) that the pressure tensor depends only on the flux factor and perhaps the energy density, (2) that the pressure tensor is prolate, (3) that the pressure in the direction of the flux matches the largest eigenvalue of the pressure tensor, and (4) that the third moment can be closed using the same functional form as the pressure tensor.

\subsubsection{The pressure tensor depends only on the flux.}
\begin{figure}
\includegraphics[width=\linewidth]{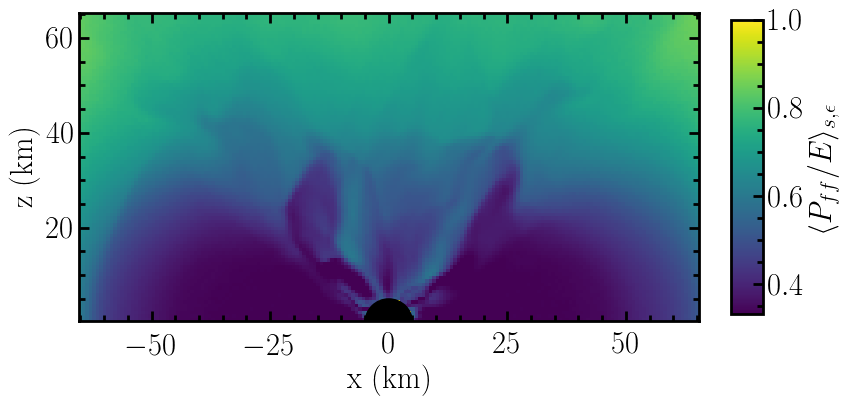}
\includegraphics[width=\linewidth]{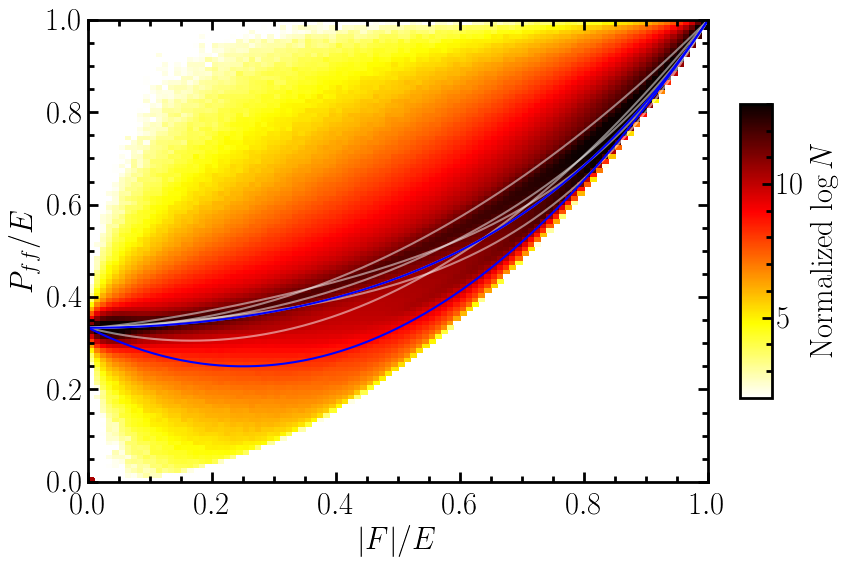}
\caption{Eddington factor. \textit{Top panel:} Neutrino pressure in the flux direction normalized by energy density on a $x-z$ slice. As expected, there is a correlation between the flux factor in Figure~\ref{fig:e_fluxfac}. \textit{Bottom panel:} histogram of the number of spatial-species-energy grid cells that have each combination of flux factor (x-axis) and Eddington factor (y-axis). The white curves show the closure relations listed in Table~\ref{tab:closure_error}. The blue curves are the MEFD maximal packing (lower) an MEFD classical (upper) closures. No simple closure can describe all grid cells.}
\label{fig:pressure}
\end{figure}
The most well-known feature of moment closures is the parameterization of the shape of higher moments based on a single quantity - the flux factor. It is also well-known that there is no single functional form of the closure that works well in all test cases (e.g., \cite{Murchikova2017}). Comparing the Eddington factor in the top panel of Figure~\ref{fig:pressure} to the flux factor in the bottom panel of Figure~\ref{fig:e_fluxfac}, there indeed appears to be a correlation between the flux factor and the magnitude of the pressure in the direction of the flux. The bottom panel of Figure~\ref{fig:pressure} shows the functional form of each of the closures in Table~\ref{tab:closure_error} on top of a histogram of the corresponding relationship between fluxes and pressures in the simulation domain. The color indicates the number of grid cells that have a flux factor and $P_{ff}$ indicated by the location on the plot. The sharp boundary on the lower right side of the colored region is a geometric limitation - it is not possible to simultaneously have all energy moving in one direction (flux factor of 1) and no pressure in that direction.

Although most closures lie close to the dark ridge, the distribution is too broad to be described by a single curve. In fact, there appears to be a second dark ridge at flux factors $\gtrsim 0.5$ near the MEFDmp curve (bottom blue curve). The bottom ridge becomes more prevalent at higher latitudes, while the main ridge is more prevalent near the equator. The majority of the closures (white and blue curves) trace the main ridge, since the equatorial regions have a smoother transition from trapped to free-streaming regimes that is more typical of spherical problems. The match between the boundary of this region and the MEFDmp curve hints that the full MEFD closure may be able to account for both ridges, but we will see in Section~\ref{sec:MEFD_results} that the extra information provided by the MEFD closure does not account for the spread.

\subsubsection{$P_{ff}$ is the largest eigenvalue.}
\label{sec:eigenvalue}
\noindent
\begin{figure}
    \centering
    \includegraphics[width=\linewidth]{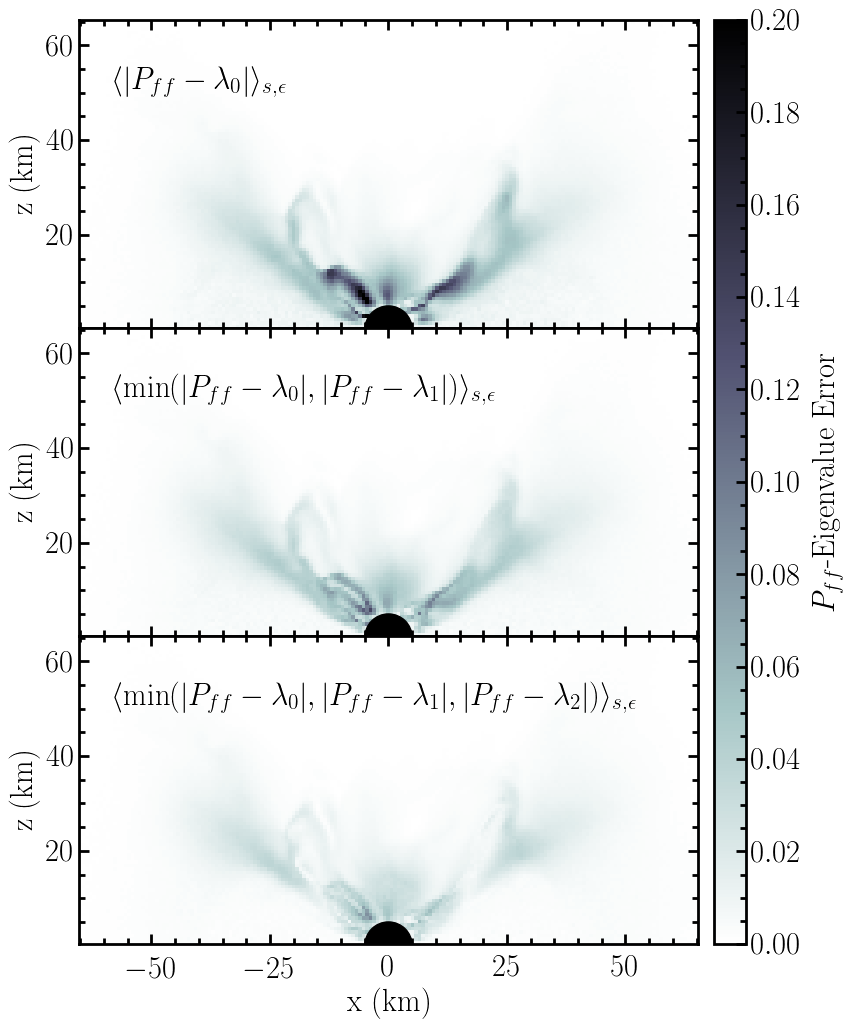}
    \caption{Misalignment between the pressure tensor and the eigenvectors. \textit{Top panel:} the energy- and species-averaged difference between the Eddington factor and the largest eigenvalue $\lambda_0$ of the pressure tensor. \textit{Middle panel:} Similar to the above, showing the energy- and species-averaged minimum of the difference between the Eddington factor and either the largest or smallest eigenvalues of the pressure tensor $\lambda_0$ and $\lambda_1$. \textit{Bottom panel:} Similar to the above, but minimizing over the difference between the Eddington factor and any of the three eigenvalues. In most regions the largest axis of the pressure tensor is parallel to the flux. Where the largest deviations from this occur, the smallest axis of the pressure tensor is largely parallel to the flux. In the interface between the disk and polar region, the flux is not well-aligned with any of the pressure tensor axes.}
    \label{fig:eigenvalue_error}
\end{figure}

The form of Equation~\ref{eq:closure_interpolation} indicates that the flux is always an eigenvector of the pressure tensor under the analytic closure approximation, since it is an eigenvector of both the diffusive and free-streaming limits. Furthermore, $P_{ff}$ must correspond to the largest eigenvalue for all but the Wilson and MEFD closures. For these closures, the Eddington factor is allowed to drop below $1/3$ at low flux factors, so $P_{ff}$ then corresponds to either the largest or smallest eigenvalue. 

The top panel of Figure~\ref{fig:eigenvalue_error} shows the energy-averaged difference between the pressure in the direction of the flux and the largest eigenvalue. For all but the Wilson and MEFD cloures, the darkenss of the color effectively indicates the amount of deviation from the closure approximation. The middle panel shows the average difference between $P_{ff}$ and either the largest or the smallest eigenvalue, so this panel shows the magnitude of the deviation from fundamental properties of the MEFD and Wilson closures. Finally, the bottom plot shows the average minimum difference from any of the three eigenvalues, and so a dark color indicates that the direction of the flux is misaligned with all of the eigenvectors, or that the orientation of the pressure tensor is weakly tied to the flux direction. 

One can also compare the pressures in other directions with the eigenvalues, though we do not show the results here. The only other local vector quantity to compare to is the three-velocity, so we can define the direction $w$ as the direction in the $F-v$ plane orthogonal to $F$, and the direction $q$ as the direction orthogonal to both. Near the equator beyond a radius of $\sim 50\,\mathrm{km}$, $P_{ff}$ matches the largest eigenvalue, $P_{qq}$ matches the smallest, and $P_\mathrm{ww}$ matches the middle one. The radiation there is moving predominantly radially, hence the dominant component of the pressure in the direction of the flux. The disk is larger in the azimuthal direction than in the polar direction, and the larger solid angle of emitting surface presented in one direction results in a larger pressure in that direction. The regions above the disk are more difficult to understand. In the diagonal regions above the disk but below the curling flow, the largest eigenvalue is best represented by $P_{qq}$, and in the curling flow $P_{ww}$ and $P_{qq}$ are both close to the largest eigenvalue, resulting in the dark regions in the top panel of Figure~\ref{fig:eigenvalue_error}. In some regions $P_{ff}$ matches the smallest eigenvalue (seen in the removal of dark regions between the top and middle panels) or the middle eigenvalue (seen as the removal of dark regions between the middle and bottom panels). We saw no clear trend in local variables that could account for all of this behavior.

\subsubsection{The pressure tensor is prolate}
\begin{figure}
    \centering
    \includegraphics[width=\linewidth]{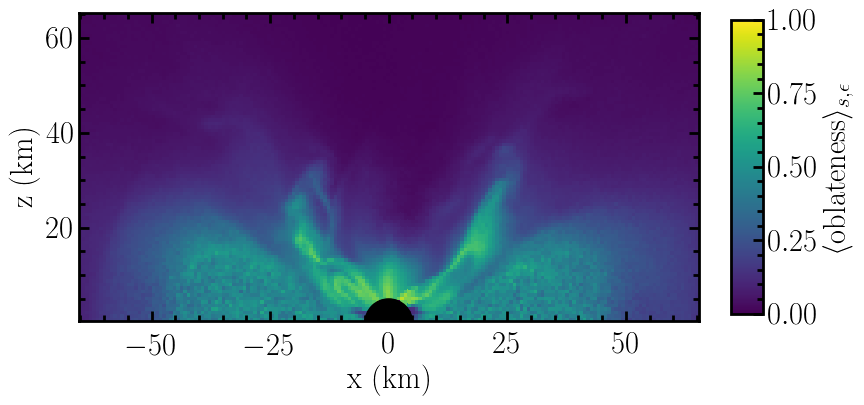}
    \includegraphics[width=\linewidth]{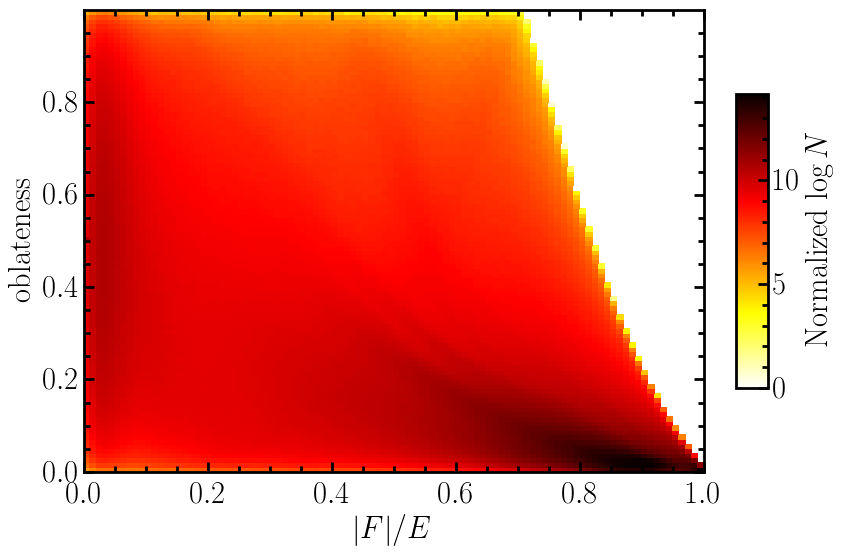}
    \caption{Oblateness of the pressure tensor as defined in Equation~\ref{eq:anisotropy_oblateness}. The analytic closures in Table~\ref{tab:closure_error} only allow for an oblateness of 0 or 1. \textit{Top panel:} $x-z$ slice showing the species- and energy-averaged oblateness. The interface between disk and polar region and the region just above the black hole show large deviations from analytic closure assumptions. The noise in the disk is because at small anisotropies, small changes in the radiation field map to radically different oblateness. \textit{Bottom panel:} histogram showing the number of species-energy-spatial grid cells with each combination of oblateness and flux factor. There is an inverse correlation between flux factor and oblateness, and fiew zones have an oblateness of zero. The dark region on the left is due to the optically-deep regions of the disk, where the flux factor is small and any oblateness can be realized through statistical Monte Carlo noise.}
    \label{fig:oblateness}
\end{figure}
Once again, for all but the MEFD and Wilson closures, the assumed pressure tensor is prolate (oblateness of zero), and for the MEFD and Wilson closures the pressure tensor can only be prolate (oblateness of zero) or oblate (oblateness of 1). The top panel of Figure~\ref{fig:oblateness} shows a $x-z$ slice of the species- and energy-averaged oblateness of the pressure tensor. Far from the disk, the oblateness indeed tends to zero as expected. In the optically deep regions of the disk, there is significant noise, since the distribution is nearly isotropic in the fluid rest frame, and small Monte Carlo statistical fluctuations correspond to large changes in oblateness because the anisotropy is so small. The first interesting note is that very close to the black hole and in the interface between the disk and polar region the oblateness can take on the full range of values. Once again, this is another indicator that the analytic closure approximation is poor in these regions. Interestingly, the oblateness is also nonzero in the equatorial optically-thin regions at radii larger than $50\,\mathrm{km}$. This is a result of the aspect ratio of the disk. As described in Section~\ref{sec:eigenvalue}, the pressure in the azimuthal direction is larger than in the $z$-direction because the disk is larger in that direction. As such, the radial pressure is largest, followed by the azimuthal pressure, followed by the $z$ pressure. The triaxial nature of the pressure tensor yields oblateness values of $\lesssim 0.5$.

The bottom panel of Figure~\ref{fig:oblateness} shows a histogram of oblateness and flux factor. The boundary at the right side of the plot is a geometric limit - in the limit of flux factor approaching 1, all of the energy must be moving in one direction and so the pressure in directions orthogonal to the flux direction tend to zero. The histogram shows that there is an inverse correlation between the flux factor and the oblateness, but once again does not follow a simple functional form. In addition, this trend varies with polar angle. The dark region on the left side of the plot comes from equatorial regions at $\cos\theta\lesssim0.5$, since this comes from the low flux factor region in the optically-deep part of the disk with near random oblatenesses. On the right side of the plot, the slope of the dark ridge is steeper at low polar angles ($\cos\theta\lesssim0.2$), indicating that some of this trend comes from the distal equatorial regions. If we only include grid zones at high polar angles ($\cos\theta\gtrsim0.6$) the slope of the dark region at $|F|/E\gtrsim 0.6$ becomes shallower than in Figure~\ref{fig:oblateness}, but then quickly rises to an oblateness of 1 at a flux factor of 0.5. This is also a geometric effect. The flux factor is small and the oblateness large at small radii despite a very low optical depth, since the radiation is crossing largely in the equatorial direction with little upward component. This trend is potentially useful information for designing a closure that uses the coordinate position as extra information for the closure, though we were unsuccessful in finding a means to do so (see Section~\ref{sec:closure_comparison}).

\subsubsection{$L$ can be closed with $\chi_p$}
\label{sec:results_L}
\begin{figure}
    \includegraphics[width=\linewidth]{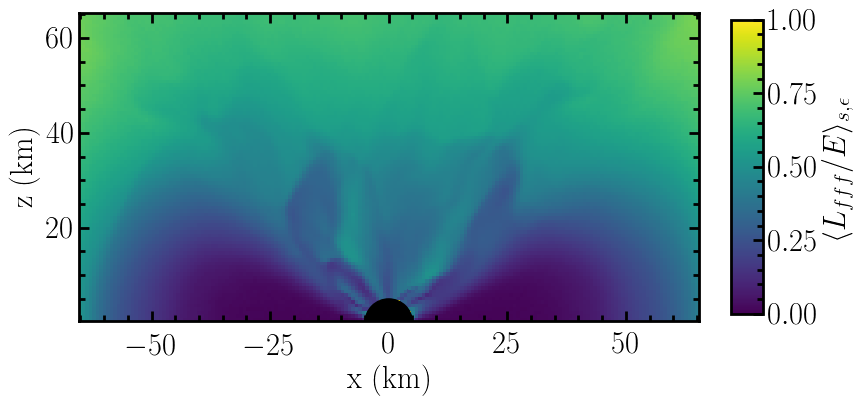}
    \includegraphics[width=\linewidth]{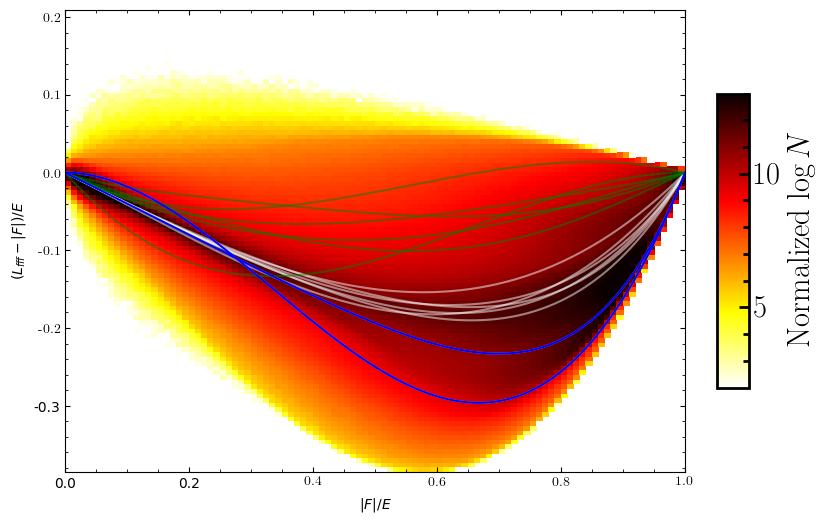}
    \caption{\textit{Top panel:} $x-z$ slice of the species- and energy-averaged component of the rank-3 moment tensor in the direction of the flux. Similar to the other moments, there is a correlation with the flux factor in Figure~\ref{fig:e_fluxfac}. \textit{Bottom panel:} histogram showing the number of cells with each combination of $L_{fff}$ and flux factor. The flux factor is subtracted from the y-axis to be able to show more detail in the plot. The white curves show the closures listed in Table~\ref{tab:closure_error}, while the blue curves show the MEFD maximal packing (lower) and MEFD classical (upper) derived in Section~\ref{sec:analytic_MEFD}. The green curves are equivalent to the white curves, but interpolating using Equation~\ref{eq:Lfree_BAD}. Most of the closures do a decent job of tracking the most dense region on the plot, but the MEFD closure covers a larger portion of the high-density region.}
    \label{fig:Lfff}
\end{figure}

The top panel of Figure~\ref{fig:Lfff} shows a slice of the component of the rank-3 tensor in the direction of the flux $L_{fff}$. Once again, there is a correlation between this and the flux factor in Figure~\ref{fig:e_fluxfac}. Since both the rank-2 and rank-3 tensors are being interpolated from optically thin to thick limits, the same closure relation is often used for both. That is, $\chi_l$ is assumed to be equal to $\chi_p$. The only closure with a self-consistent third moment interpolator is the MEFD closure, which we derive in Section~\ref{sec:analytic_MEFD}.

The bottom panel of Figure~\ref{fig:Lfff} shows a histogram of the flux factors and values for $L_{fff}$, with the flux factor also subtracted from the $y$-axis value to show more detail in the plot. Similar to the case of the pressure histogram in Figure~\ref{fig:pressure}, there is an upper and a lower dark ridge and the distribution is too broad to be described by a single curve. The white curves show values of $L_{fff}$ inferred by using the $\chi_p$ from the bottom set of closures in Table~\ref{tab:closure_error} in lieu of an appropriately derived $\chi_l$. The blue curves show the results from the $\chi_l$ derived in Section~\ref{sec:analytic_MEFD} for the MEFD maximal packing (lower) and classical (upper) limits. It seems that the white curves do generally follow the upper ridge, but are unable to account for the lower ridge. The MEFD maximum packing curve, however, once again nicely encompasses this region, leading to a hope that the MEFD closure may be able to account for both regions. However, we will show in Section~\ref{sec:MEFD_results} that in this snapshot the MEFD closure closely resembles its classical limit and the extra information from the degeneracy cannot explain the spread.

The green curves in Figure~\ref{fig:Lfff} show the results when using Equation~\ref{eq:Lfree_BAD} for the free-streaming limit. This results in unequivocally poor results. Although one can construct an interpolator for this flavor of interpolation (e.g., Equation~\ref{eq:chil_BAD} for the MEFD closure), it is more straightforward to use Equation~\ref{eq:PL_diff_free}. Doing so requires that the trace of the rank-3 tensor is the flux and makes $\chi_p$ a reasonable approximation for $\chi_l$.

\subsection{Specific Closures}
\label{sec:specific_closures}
The particular closures we compare to are listed in Table~\ref{tab:closure_error} and shown in the bottom panels of Figures~\ref{fig:pressure} and \ref{fig:Lfff}. The Thick and Thin closures simply take the corresponding limit in Equation~\ref{eq:closure_interpolation} irrespective of the flux factor, providing a sense of scale for the errors the other more reasonable closures make. The MEFD closure along with it's classical (MEFDc) and maximum-packing (MEFDmp) limits were described in detail in Section~\ref{sec:analytic_MEFD} and guarantee that the set of moments is possible to realize with a fermionic radiation field. This is also the only closure with a self-consistent closure for the third moment. The Levermore closure \cite{Levermore1984} is derived by assuming that the radiation field is isotropic in the frame where the net radiation flux is zero. The Kershaw {closure} \cite{Kershaw1976} {is just a simple, non-unique quadratic function interpolating between the optically thick and thin limits in a way that is always realizable for a Bose-Einstein gas. The Wilson \cite{Wilson1975} closure is the harmonic mean of the diffusive and free-streaming limits \cite{Smit2000}. Finally, the Janka closures \cite{Janka1991} were determined from fits to Monte Carlo neutrino transport data in one-dimensional simulations of core-collapse supernovae.

\subsubsection{The MEFD Closure}
\label{sec:MEFD_results}
\begin{figure}
    \centering
    \includegraphics[width=\linewidth]{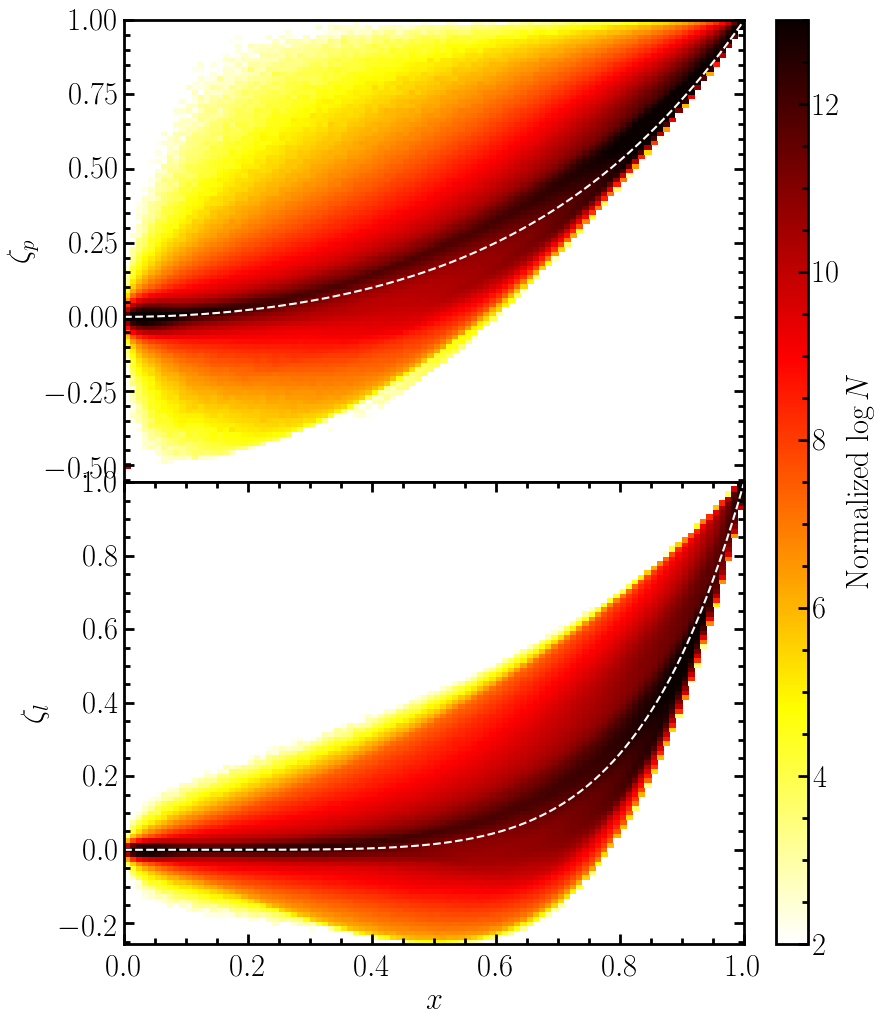}
    \caption{Evaluation of the maximum entropy closure assumptions. \textit{Top panel:} 2-D histogram of the flux saturation $x$ and the maximum-entropy universal pressure-closure curve (Equation~\ref{eq:MEFD_approximate}). \textit{Bottom panel:} Similar, but for the semi-universal rank-3 moment closure curve (Equation~\ref{eq:MEFD_approximate}). If the Monte-Carlo derived distributions looked like the maximum-entropy distributions (Equation~\ref{eq:f_MEFD}), all points would lie along the dotted white curve in the top panel. The curve is not as universal away from the $x=\{0,1\}$ limits, but this shows that the approximate curve neatly lies within the most dense regions of the plot.}
    \label{fig:MEFD_closure_fit}
\end{figure}
Since the MEFD maximum packing curve neatly outlines the bulk of the dark regions in Figures~\ref{fig:pressure} and \ref{fig:Lfff}. Unlike any of the other closures, the MEFD closure also uses the occupation number as input, so one might be tempted to guess that the spread is neatly accounted for by this extra information. The salient feature of the MEFD closure for the pressure tensor is that there is a single universal curve of $\zeta_p(x)$ (Equation~\ref{eq:zeta_0}) for all values of the occupation number, and the effects of the occupation number come in only through Equation~\ref{eq:MEFD_closure} and the definition of the flux saturation $x$. The top panel of Figure~\ref{fig:MEFD_closure_fit} is similar to the bottom panel of Figure~\ref{fig:pressure}, though the x-axis is flux saturation $x$ instead of flux factor, and the y-axis is pressure saturation $\zeta_p$ instead of Eddington factor. The white curve shows the approximate universal function from Equation~\ref{eq:MEFD_approximate}. If the information from the occupation number were able to account for the spread of the dark regions in Figure~\ref{fig:pressure}, we would expect the dark regions to collapse to the white line in this figure. Unfortunately, that is not the case - since the occupation numbers only reach at most $0.5$, $\{|F|/E,P_{ff}\}$ and $\{x,\zeta_p\}$ are nearly identical. The same applies to the bottom panel, which shows the rank-3 saturation $\zeta_l$ and the semi-universal curve in Equation~\ref{eq:MEFD_approximate}. Since $\zeta_l$ is not universal except in the limits of $x\rightarrow\{0,1\}$, we do not expect the distribution to collapse to a single line, but this does comfortingly demonstrate that the approximate curve does lie in the most dense regions of the plot.

\subsubsection{Other Closures}
\label{sec:closure_comparison}
\begin{table}
\begin{tabular}{lcccccc}
Closure & $P_{ff}/E$ & $\Theta$ & A & $L_{fff}/E$ & $L_4/E^2$ & $\alpha\mathcal{F}^\mu_\mathrm{ann}u_\mu$\\
 & $\times100$ & $\times100$ & $\times100$ & $\times100$ & $\times100$ & $\times100$ \\\hline
Thick & 12.6 & -- & 29.8 & 2.8 & 1.5 & {-16.6} \\
Thin & 17.7 & 6.9 & 35.6 & 2.15 & 1.93 & {87.4} \\\hline
MEFD & 0.297 & 6.9 & 1.20 & {0.316} & {0.33} & {1.01} \\
MEFDc & 0.297 & 6.9 & 1.20 & {0.316} & {0.33} & {1.10} \\
MEFDmp & 0.939 & 13.6 & 2.26 & {0.705} & {0.42} & {-4.76} \\\hline
Levermore & 0.233 & 6.9 & 0.93 & 0.231 & 0.22 & {3.98} \\
Kershaw & 0.32 & 6.9 & 0.90 & 0.339 & 0.28 & {11.1} \\
Wilson & 0.33 & 10 & 1.27 & 0.237 & 0.23 & {2.17} \\
Janka1 & 0.442 & 6.9 & 1.59 & 0.198 & 0.22 & {-1.76} \\
Janka2 & 0.274 & 6.9 & 0.96 & 0.213 & 0.21 & {2.48} \\
\end{tabular}
\caption{Integrated difference between the Monte Carlo and indicated closure solution for representative components of rank-2 and rank-3 moment tensors. Numbers are displayed to the first digit that changes using a similar calculation with 0.4 times as many Monte Carlo particles and are multiplied by 100 to remove leading zeros for display. $P_{ff}$ and $L_{fff}$ are the component of the neutrino pressure tensor and rank-3 tensor along the direction of the flux. $\Theta$ is the oblateness and $A$ is the anisotropy (Equation~\ref{eq:anisotropy_oblateness}). $L_4$ is the scalar invariant of the rank-3 tensor (Equation~\ref{eq:L4}). $\alpha\mathcal{F}^\mu_\mathrm{ann}u_\mu$ is the rate of increase of the thermal energy by neutrino pair processes as measured by an observer at infinity. The pair process error is an actual error rather than a $\chi^2$ value, comparing to the Monte Carlo result of $4.78\times10^{50}\,\mathrm{erg\,s}^{-1}$}
\label{tab:closure_error}
\end{table}

Table~\ref{tab:closure_error} shows the errors in integrated values of various quantities relative to the Monte Carlo results. Specifically, the numbers for a quantity $q$ are computed as
\begin{equation}
\frac{1}{n_x n_y n_z n_\epsilon} \sum_{i,j,k,l} (q_{\mathrm{closure},i,j,k,l} - q_{\mathrm{MC},i,j,k,l})^2\,\,,
\end{equation}
where the prefactor contains the number of grid zones in $x$, $y$, $z$, and energy, and the sum is over the corresponding grid zones labeled by $\{i,j,i,l\}$. The exception is the farthest right column, which we discuss in Section~\ref{sec:annihilation_results}

The thick and thin closures are obviously poor choices, but are shown for reference of scale. Many of the closures are similarly accurate, since they all generally lie within the rather broad distribution of flux factor and pressure/rank-3 tensor in Figures~\ref{fig:pressure} and \ref{fig:Lfff}. Even so, the Janka 2 closure shows the smallest error for $P_{ff}/E$  and $L_4/E^2$ and the Janka 1 closure for $L_{fff}/E$. These are followed closely by the MEFD closure, which performs reasonably well in the pressure categories (columns 2-4), although somewhat poorly in the rank-3 categories (columns 5-6). Ironically, using the rank-2 closure for the rank-3 moments produces smaller errors than the rank-3 closure freshly derived in Section~\ref{sec:analytic_MEFD}}. 

Many of the closures yield similar errors for the oblateness $\Theta$ because the Thin, MEFDc, Levermore, Kershaw, and Janka closures all assume an oblateness of exactly 1. The MEFD, MEFDmp, and Wilson closures do allow for an oblateness of exactly 1 at low flux factors (as indicated by the curves dipping below 1/3 in Figure~\ref{fig:pressure}), resulting in a larger error. The only way to drive this error smaller is to create a closure that allows for triaxial pressure tensors. We attempted to create such a closure by assuming the oblateness follows $\Theta=(1-|F|/E)^2$ (estimated from Figure~\ref{fig:oblateness}), setting the eigenvector with the largest eigenvalue along $F/E$, that with the smallest eigenvalue along the component of the three-velocity orthogonal to $F/E$, and that with the middle eigenvalue along the direction perpendicular to both. This results in a smaller oblateness error of $0.0442$, though at the cost of a marginal increase of the error in $P_{ff}/E$ to 0.00304. We were unable to find a good way to set the oblateness and orientation of the pressure tensor using only local variables, since the trends differ in different regions of the system (see Section~\ref{sec:assessing_assumptions}). In addition, although the MEFD and Levermore closures guarantee a realizable distribution (i.e. they never require occupation numbers larger than 1 or smaller than 0), once we break the assumptions on the symmetry directions used in the derivation, it is not clear how to ensure that the triaxial closure is realizable.

It is worth noting that the MEFD and MEFDc closure yield nearly identical results, indicating that the neutrino field in this snapshot is not very degenerage (see Figure~\ref{fig:e_fluxfac}). The ability of the MEFD closure to yield realizable moments in highly degenerate scenarios is little-used. For the same reason, the MEFDmp closure yields comparatively large errors, since the closure assumes that the distribution is maximally degenerate for a given flux factor.

\subsection{Annihilation}
\label{sec:annihilation_results}
\begin{figure}
    \centering
    \includegraphics[width=\linewidth]{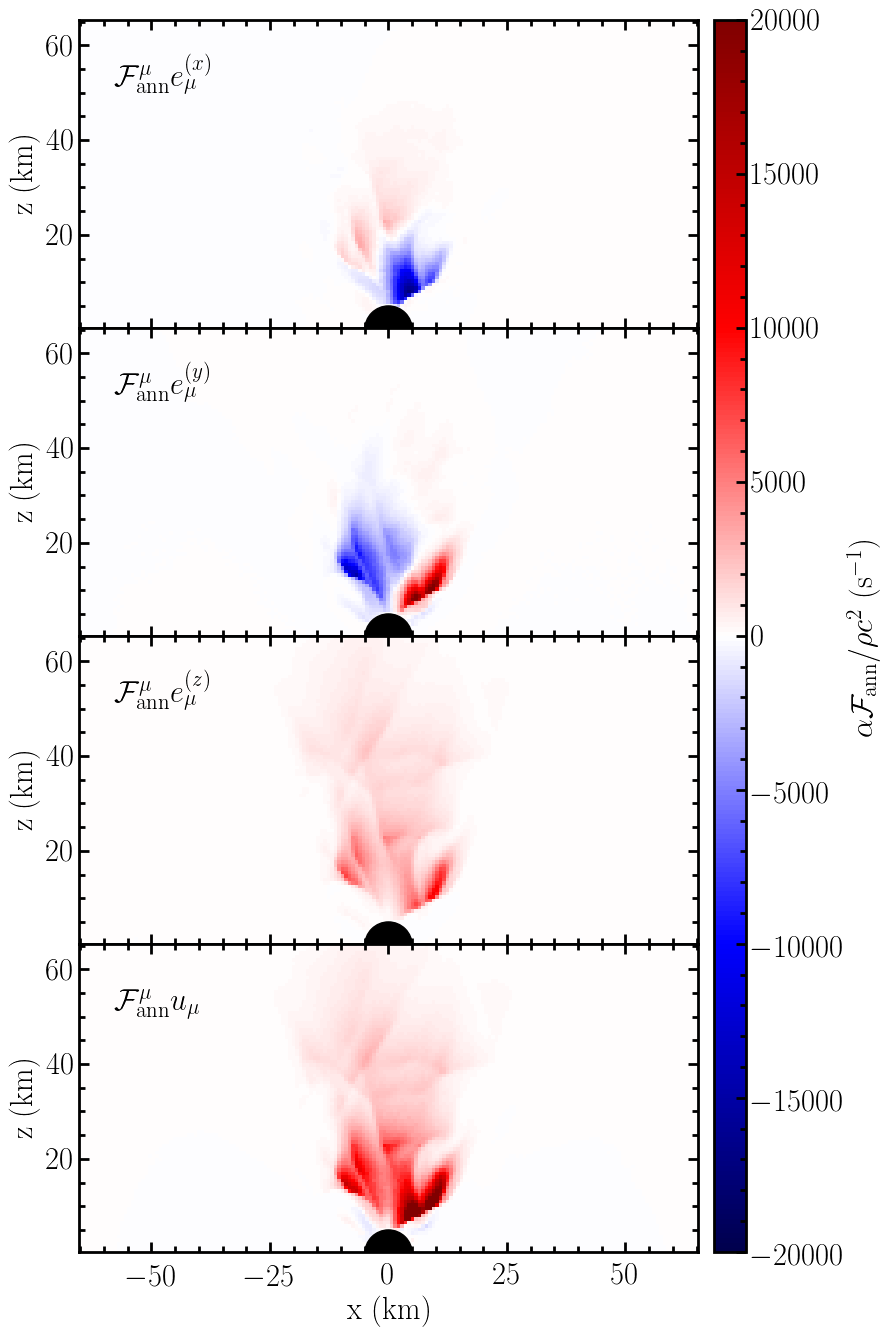}
    \caption{Neutrino pair annihilation rate normalized by mass-energy density. The rate of deposition of $x$-, $y$-, and $z$-momentum are shown in the first, second, and third panels, respectively. The rate of thermal energy deposition is shown in the bottom panel. Neutrino pair annihilation is not dynamically important in the disk, but in the polar regions it rapidly heats the low-density matter and drives it inward, around in the direction of the disk's orbit, and upward.}
    \label{fig:annihilation}
\end{figure}

Figure~\ref{fig:annihilation} shows the momentum (top three panels) and thermal energy (bottom panel) deposition rate in a slice of the domain normalized by the mass energy density. Since we include both absorption and emission, the net effect on the dense optically-thick regions of the disk is minimal. The value of the number then indicates the rate at which the thermal energy density or momentum becomes relativistic. The top panel shows that the right side of the polar region is being pushed left, and the left side right, possibly helping to collimate the flow very close to the black hole. The second panel shows the azimuthal momentum deposition, which is in the direction of the disk orbit. The third panel shows the $z$-component of the momentum deposition, indicating that the annihilation can directly provide a great deal of upward momentum. Finally, the bottom panel shows the rate of deposition of thermal energy. Even without direct momentum, this thermal energy will drive a polar outflow.

For an order-of-magnitude estimate of the power available from neutrino annihilation for powering polar outflows, we compute the net deposition of thermal energy in the region within 45 degrees of the polar axis. The fiducial Monte Carlo annihilation calculation yields $4.78\times10^{50}\,\mathrm{erg/s}$ of deposited thermal energy. The specific heating rate, annihilation power, and density in the polar region are comparable to that seen in dynamical calculations including neutrino annihilation \cite{Just2016,Fujibayashi2017}. Because of this, we echo the conclusions of these works that the neutrino pair annihilation will modify the dynamics and leptonization of polar ejecta, but the mass in this region will likely preclude a neutrino-driven jet.

The time component of the four-force due to neutrino pair annihilation in Equation~\ref{eq:annihil_fourforce} only depends on the number of neutrino field moments equal to the number of terms used in the Legendre expansion of the kernel $\Phi$. Using this, we can demonstrate the importance of each of these terms. If we only include the $\Phi_0$ term, the annihilation power comes out to $1.22\times10^{51}\,\mathrm{erg\,s}^{-1}$ (larger by a factor of 2.5 than the fiducial result above using three terms). Including only the $\Phi_0$ and $\Phi_1$ terms yields $3.98\times10^{50}\,\mathrm{erg\,s}^{-1}$, or 0.83 times the fiducial result. Thus, including the estimate for the third moment of the annihilation kernel in Equation~\ref{eq:phiannihil2} has roughly a $17\%$ effect on the available annihilation power.

Since the term in Equation~\ref{eq:annihil_fourforce} with $\Phi_2$ depends on the pressure tensor, the choice of closure can affect the calculated annihilation rate in a two-moment radiation transport scheme. The farthest right column of Table~\ref{tab:closure_error} shows the relative error of the integrated annihilation power due to the choice of closure. The MEFD and MEFDc closures exhibit the smallest error, while the Kershaw closure yields a large error of 1{1}\%. The rest of the other closures yield an error of only a few percent. Thus, despite the complexity of the radiation field in the polar regions, the choice of a closure is important only to get percent-level accuracy of the integrated annihilation power.

Finally, we briefly note the impact of two other assumptions. If we do not subtract off the mass energy of the electron-positron pairs, we get a polar annihilation power of $5.85\times10^{50}\,\mathrm{erg\,s}^{-1}$ (difference of 22\%). Second, although we include the  $\Phi^{(p)}$ terms in Equation~\ref{eq:annihil_fourforce}, they do not actually affect the polar annihilation power to the presented accuracy, since the vast majority of the neutrino pair production is occurring in the dense disk.

\section{Conclusions}
\label{sec:conclusions}
We use a newly developed steady-state Monte Carlo radiation transport code to evaluate assumptions used by analytic closures to the moment equations for radiation transport in neutron star merger disks.

We first extend the MEFD closure to include the rank-3 moment for use in spectral M1 simulations (Figure~\ref{fig:MEFD_l}). The proposed approximation to the closure (Equation ~\ref{eq:MEFD_approximate}) accurately represents the full solution under the assumptions that go into the closure to at most 3.5\% error for all flux factors and neutrino degeneracies. This is the the only closure with a self-consistent treatment of the third moment, though we do not expect the impact to be large.

In order to test this and other closures in the context of neutron star merger simulations, we developed the Monte Carlo neutrino transport code {\tt SedonuGR} to compute the steady-state neutrino radiation field on a static discretized fluid and spacetime background from a three-dimensional simulation snapshot in three dimensions. We calculate the full steady-state radiation field including moments up to rank-3 and compare these moments against fundamental assumptions used in all analytic closure relations (Section~\ref{sec:assessing_assumptions}). We demonstrate the expected result that a single analytic closure is unable to reproduce the Eddington factor (Figure~\ref{fig:pressure}). Furthermore, the largest axis of the pressure tensor ellipsoid is not aligned with the flux direction just above the black hole and in the interface between the disk and the evacuated polar regions (Figure~\ref{fig:eigenvalue_error}), contrary to the assumptions commonly used in generating analytic closures. In these same regions, the pressure tensor ellipsoid is largely triaxial (Figure~\ref{fig:oblateness}), unlike the prolate ellipsoid assumed by analytic closures. This is also true near the equator outside of the dense part of the disk due to the aspect ratio of the disk. Finally, we demonstrate that the analytic relations used to determine the pressure tensor are also reasonable closures for the rank-3 moment (Figure~\ref{fig:Lfff}), though they once again cannot explain the full spread seen in various parts of the disk.

None of the closures listed in Table~\ref{tab:closure_error} spring out as an obvious best choice, including the MEFD closure that we so carefully extend, though the MEFD, MEFDc, Levermore, and Janka 2 closures were as accurate as could be expected by such a closure. Although we tried to use additional information like the oblateness of the pressure tensor to improve the analytic closures, the lack of clear trends made improvements based on estimations of little benefit.

Finally, we briefly touch on the impact of the moment expansion in calculating the rate of deposition of energy and momentum in the polar regions (Figure~\ref{fig:annihilation}). Assuming the annihilation kernels are expanded in terms of Legendre polynomials, keeping up to the third term in the expansion (which involves the pressure tensor) yields a $17\%$ enhancement of the net annihilation power over keeping only the first two terms. The choice of closure among those listed in Table~\ref{tab:closure_error} made at most a $5\%$ impact on the annihilation power.

We intentionally avoid investigating non-local closures (e.g., closures that depend on the coordinate position or on derivatives of the radiation field) because they fundamentally change the nature of the transport equation. For instance, if the pressure tensor is evaluated based on the gradient of the flux factor, the flux of the neutrino flux would then depend on the second derivative of itself, adding an elliptic character to an otherwise hyperbolic equation.

It may yet be possible to construct a closure specific to neutron star mergers using, for example, a neural network and exact transport data from a large number of snapshots of many models evolved using exact methods like Monte Carlo \cite{Foucart2018,Miller2019a}, discrete ordinates \cite{Nagakura2017a}, or characteristics \cite{Davis2012,Weih2020}. In addition, it is possible to extend moment methods to dynamically evolve the pressure tensor as well (e.g., \cite{Banach2005,Banach2013}), requiring a closure for the rank-3 and rank-4 moments. However, the problem certainly appears to be complex enough to warrant using full transport methods directly.

\section{Acknowledgements}
We are grateful to David Radice for providing the 3D model of a neutron star merger accretion disk. We also thank Sanjana Curtis, Carla Fr\"ohlich, Dan Kasen, and Hiroki Nagakura for useful discussions, and Jim Kneller and Carla Fr\"ohlich for exclusive access to computing resources. SR is supported by the N3AS Fellowship under National Science Foundation grant PHY-1630782 and Heising-Simons Foundation grant 2017-228. This material is based upon work supported by the National Science Foundation under Award No. AST-2001760. Software used: NumPy \cite{VanDerWalt2011}, Matplotlib \cite{Hunter2007}, and Mathematica \cite{Mathematica}.

\bibliography{references}

\appendix

\section{Code Tests}
\label{app:code_tests}
\subsection{Schwarzschild Geodesics}
Geodesics in a Schwarzschild spacetime have simple analytic solutions, making them ideal tests for the accuracy of the general relativistic particle integrator. We construct a coarse spacetime in spherical symmetry with an inner radius of $r_\mathrm{sch}$ and grid zones spaced as $r_{\mathrm{out},i}=1.5^{1/p} r_\mathrm{sch}$. We perform the test with three resolutions using $p=1,2,4$. In all of the tests, neutrinos start at the photon sphere ($r=1.5r_\mathrm{sch}$) where a photon can orbit the black hole circularly. We try to expose the code to challenging tests that expose the limitations of a spacetime represented on a discrete grid.

\begin{figure}
    \centering
    \includegraphics[width=\linewidth]{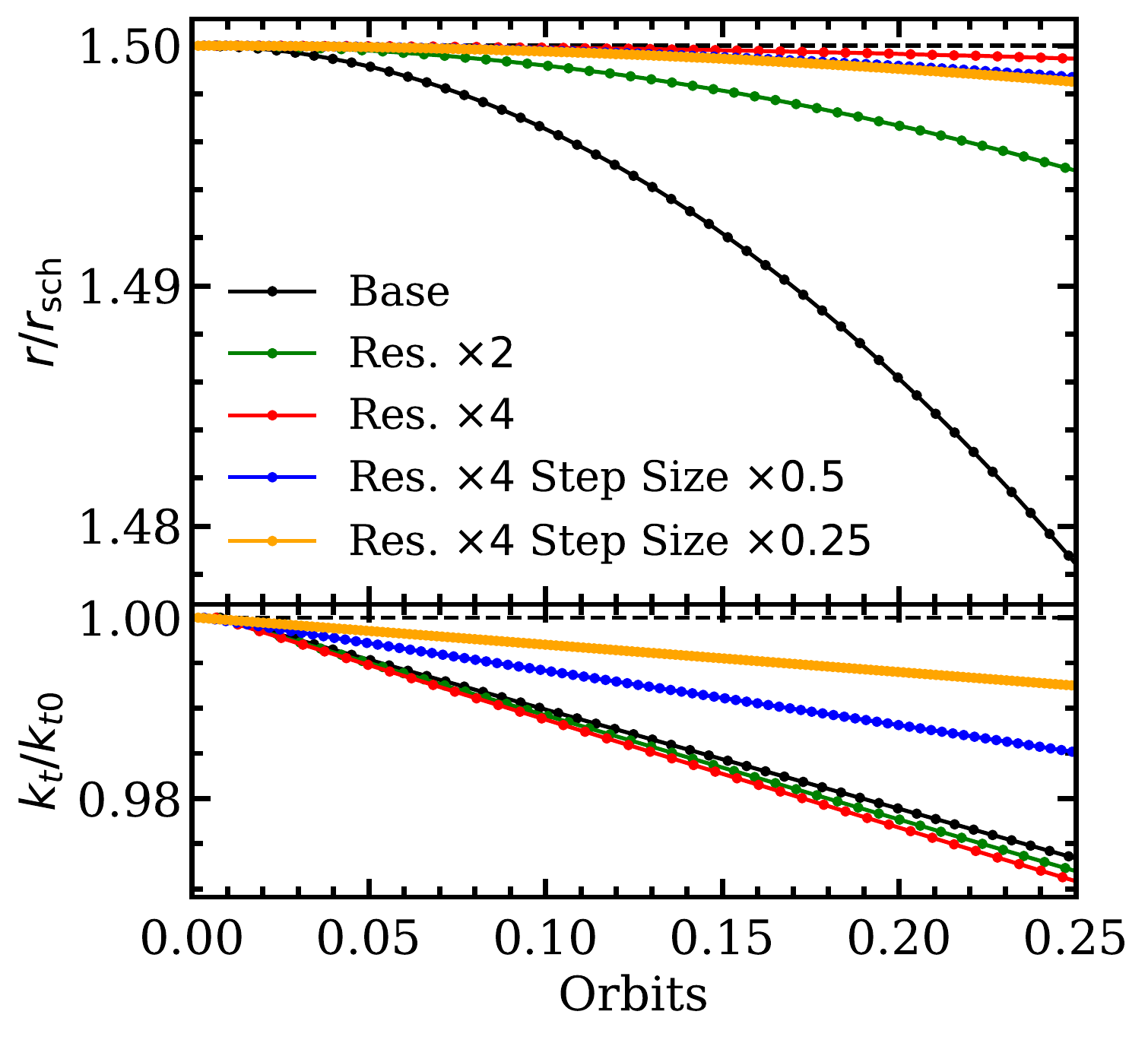}
    \caption{Neutrino orbit test. A neutrino starts at a radius of $1.5 r_\mathrm{sch}$ moving tangentially to the central black hole. The top panel shows the neutrino radius and the bottom panel shows the conserved energy (time component of $k_\mu$), both of which are expected to remain constant (dashed lines). The curve color delineates the grid resolution and particle step size.}
    \label{fig:schwarzszetald_around}
\end{figure}
In Figure~\ref{fig:schwarzszetald_around} we show trajectories of neutrinos moving azimuthally around the black hole, where we expect the radius (top panel) and the $k_t$ component of the momentum (bottom panel) to remain constant as the photon orbits the black hole (dashed lines). Since the starting radius of $1.5r_\mathrm{sch}$ is at the interface between two grid cells and the values of the metric are interpolated from the grid centers, the metric values are rather approximate elsewhere in the grid cell. The derivatives of the metric, however, are constant between two grid zone centers and are most accurate at the zone interfaces. The coarseness of the grid clearly shows, as after a quarter of an orbit the neutrino has diverged from the Schwarzszetald solution by $0.02 r_\mathrm{sch}$. As we double (green curve) and quadruple (red curve) the grid resolution while keeping the neutrino step size constant, the errors reduce, showing second order convergence. However, all three of these cases show similar errors in the conservation of the neutrino's momentum (bottom panel). If we shorten the neutrino step by a factor of two (blue curve) and four (yellow curve), the conserved momentum is better conserved, though it shows only first order convergence.

\begin{figure}
    \centering
    \includegraphics[width=\linewidth]{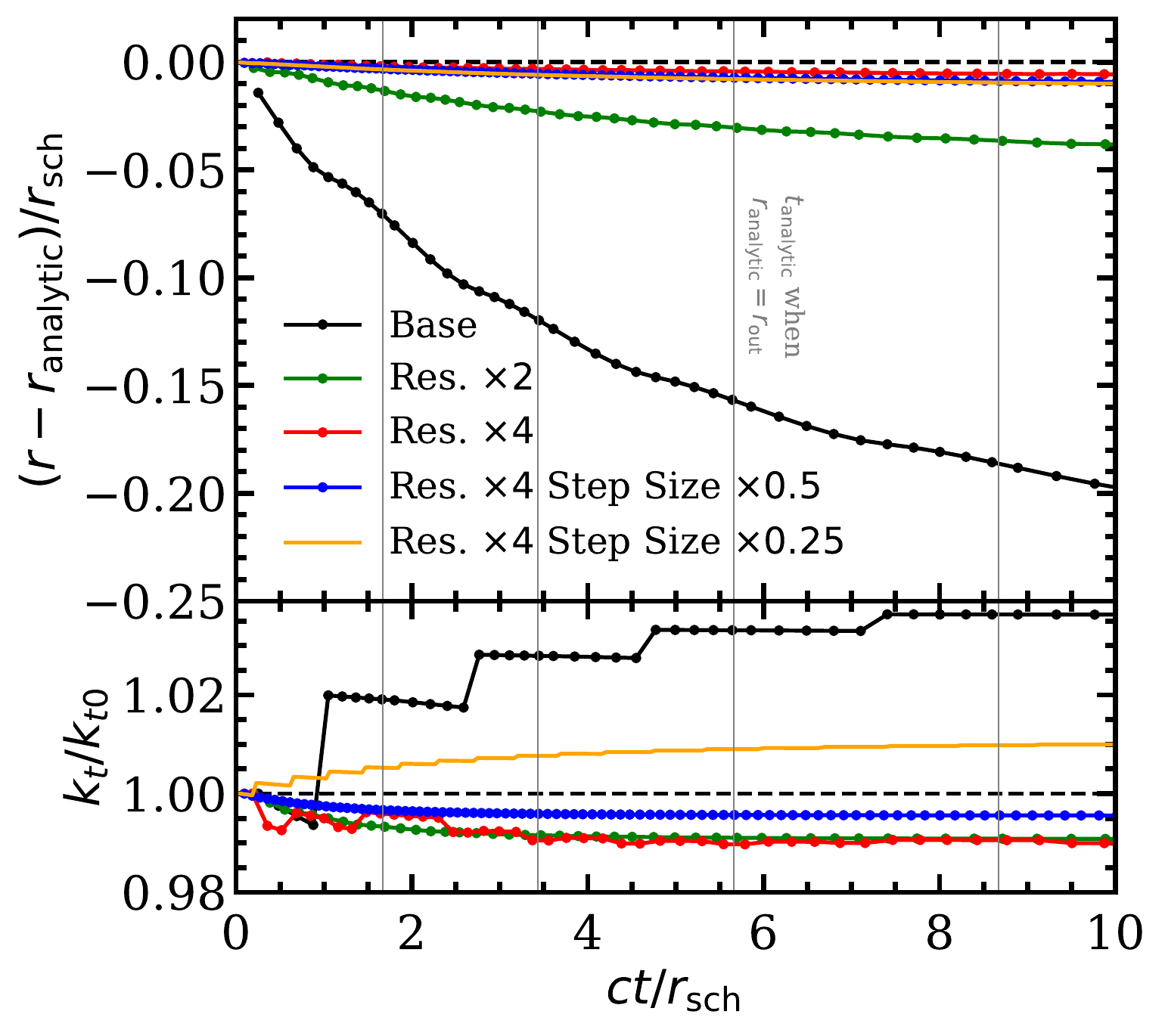}
    \caption{Neutrino radial trajectory test. A neutrino starts at a radius of $1.5r_\mathrm{sch}$ moving radially outward. The top panel shows the deviation of the neutrino radius from the analytic solution and the bottom panel shows the violation of the neutrino conserved energy. The colors of the curves differentiate the grid resolution and particle step size. The gray vertical lines show the time required for a neutrino to reach each grid cell boundary in the base-resolution grid based on the analytic solution.}
    \label{fig:schwarzszetald_radial}
\end{figure}
Figure~\ref{fig:schwarzszetald_radial} shows deviations from an analytic solution for the trajectory of a radially-moving neutrino starting at the same radius of $1.5 r_\mathrm{sch}$. The coordinate time at which the neutrino reaches radius $r$ is given by
\begin{equation}
    ct(r) = \left(r + r_\mathrm{sch}\ln\left(\frac{r}{r_\mathrm{sch}}-1\right)\right)-r_\mathrm{sch}\left(1.5 + \ln\left(0.5\right)\right)\,\,.
\end{equation}
The colors of the curve represent calculations with the same parameters as in Figure~\ref{fig:schwarzszetald_radial}. In addition, we plot the analytical time required for a neutrino to get to a grid cell boundary plotted as gray vertical lines as a proxy to show the size of individual grid cells at the base resolution. For the neutrino traversing through the base resolution grid (black curve), sizable errors build up, but we once again see second order convergence as we double (green) and quadruple (red) the grid resolution keeping the neutrino step size fixed. The conserved energy (bottom panel) shows a steady accumulation of error until the neutrino crosses the bin center, at which point the metric derivatives change discontinuously, resulting in a discontinuous jump in the error of the solution.

\subsection{Neutrino Oven}
\begin{figure}
    \centering
    \includegraphics[width=\linewidth]{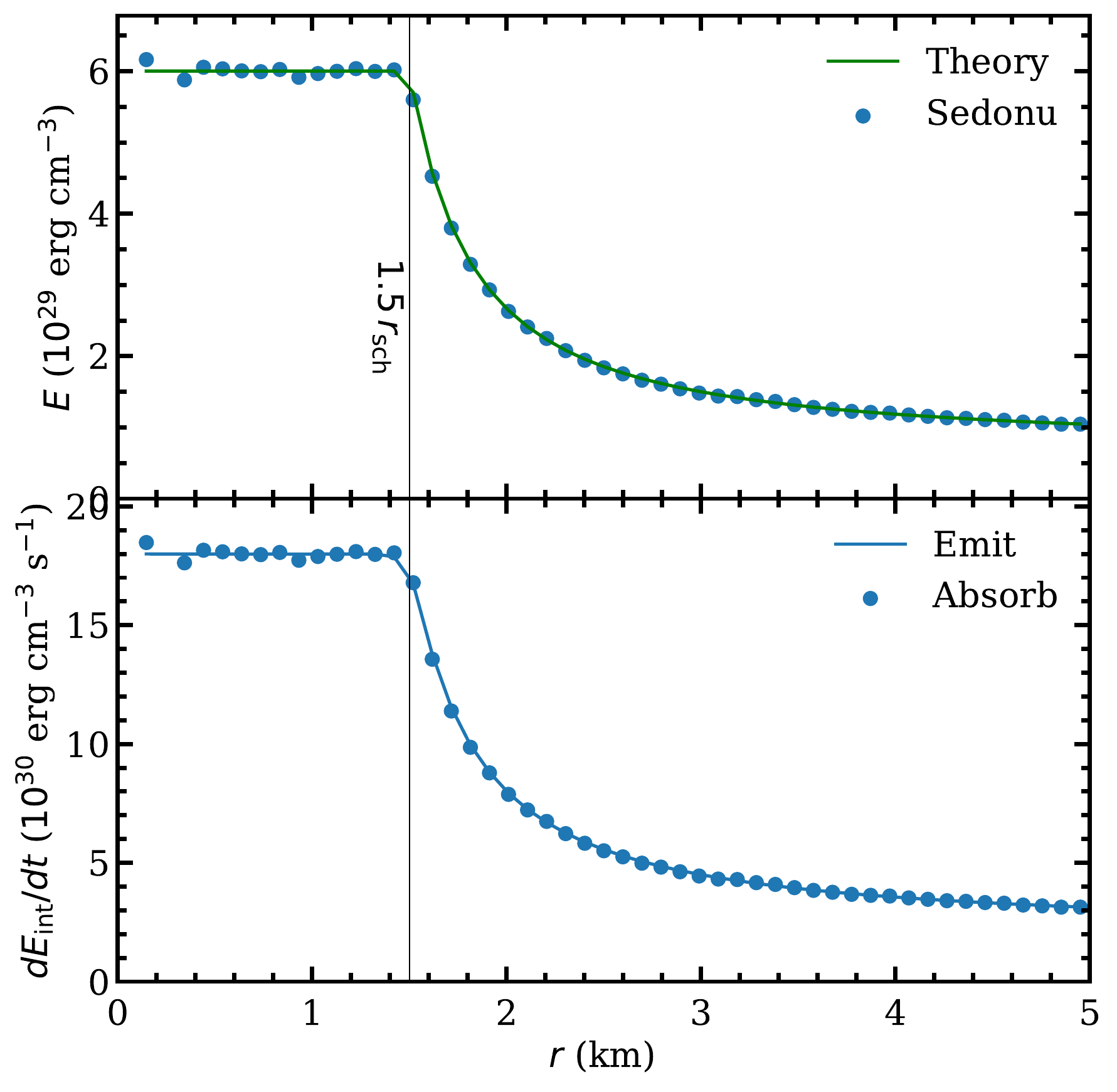}
    \caption{Neutrino oven test. Equilibrium neutrino energy density (top panel), energy emission/absorption rates (bottom panel). Noise increases toward the center due to fewer packets passing through those cells. The spacetime is a Scwarzszetald solution characterized by $r_\mathrm{sch}=1\,\mathrm{km}$ outside of $r=1.5r_\mathrm{sch}$ (vertical line), and constant inside (Equation~\ref{eq:neutrino_oven}).}
    \label{fig:neutrino_oven}
\end{figure}
Now that the errors associated with motion in a discrete spacetime are established, we move on to a full equilibrium test of the transport. We set up a spherical spacetime generated by a shell of mass at $r=1.5 r_s$. That is, our metric is
\begin{equation}
\label{eq:neutrino_oven}
    \begin{aligned}
        \alpha^2 &= 1-\frac{r_s}{\max(r,1.5 r_s)}\,\,, \\
        X &= \frac{1}{\alpha}\,\,.
    \end{aligned}
\end{equation}
We know that the equilibrium temperature should vary with $T \propto 1/\sqrt{g_{tt}}$ \cite{Tolman1930}. For concreteness, we choose $r_s=1.5\,\mathrm{km}$ and a core temperature (inside $1.5r_s$) of $10\,\mathrm{MeV}$. We set the fluid temperature outside of the core according to $T=T_\mathrm{core} \alpha_\mathrm{core}/\alpha$ and check that the radiation field, absorbed energy, and absorbed lepton number match a Blackbody distribution, the emitted energy, and the emitted lepton number, respectively. We set a gray scattering opacity of $\kappa_s=1\,\mathrm{cm}^{-1}$ and a gray absorption opacity of $\kappa_a=10^{-10}\,\mathrm{cm}^{-1}$. This ensures that each grid zone is very optically deep to scattering, forcing us to use the new random walk Monte Carlo implementation. The low absorption opacity, however, causes the effective absorption optical depth of a grid zone to be $\kappa_\mathrm{eff} \Delta r = \sqrt{\kappa_a \kappa_s}(10^4\,\mathrm{cm})=0.1$. If the absorption optical depth is too large, neutrinos never actually leave their cell of origin and the inter-cell transport properties of the random walk algorithm are not tested. We use an evenly-spaced energy grid of 300 bins from 0 to $150\,\mathrm{MeV}$. In this test we emit 10 packets from each radius-energy bin for a total of 147000 packets. We roulette the packets if their weight decreases to $10^{-3}$ of their original weight. Finally, the step size, and hence randomwalk sphere size, is set to always be $0.4\Delta r$.

Despite being a simple one-dimensional test, the calculation is actually still rather expensive due to the fact that the neutrinos scatter many times, and random walk events are significantly more expensive than regular scattering events since each step requires three separate move/interpolate events (see \ref{sec:randomwalk}). 

The results from simulating match the theoretical predictions well. The top panel of Figure~\ref{fig:neutrino_oven} shows the neutrino energy density in each grid cell (dots) and the theoretical expectation (green curve). At small radii the noise is much larger since fewer packets pass through the cells due to an equal number of packets being created in each grid zone, and the noise decreases with particle count.

We found that in this test it is particularly important to have a second-order integration of the neutrino momentum. Using a scheme that was second order in the neutrino position but first order in the neutrino momentum resulted in a over/undershoot at the cusp.

\end{document}